\documentclass[longauth]{aa} 
\usepackage{natbib}
\usepackage{fixltx2e} 
\usepackage{graphicx}
\usepackage{txfonts}
\def \agile {\textrm{AGILE}}

\begin{document}
   \title{MAGIC gamma-ray and multi-frequency observations of flat spectrum radio quasar PKS~1510$-$089 in early 2012}


%
\author{
J.~Aleksi\'c\inst{1} \and
S.~Ansoldi\inst{2} \and
L.~A.~Antonelli\inst{3} \and
P.~Antoranz\inst{4} \and
A.~Babic\inst{5} \and
P.~Bangale\inst{6} \and
U.~Barres de Almeida\inst{6} \and
J.~A.~Barrio\inst{7} \and
J.~Becerra Gonz\'alez\inst{8,}\inst{25} \and
W.~Bednarek\inst{9}
% ...
}
\institute {
IFAE, Campus UAB, E-08193 Bellaterra, Spain
\and Universit\`a di Udine, and INFN Trieste, I-33100 Udine, Italy
\and INAF National Institute for Astrophysics, I-00136 Rome, Italy
\and Universit\`a  di Siena, and INFN Pisa, I-53100 Siena, Italy
\and Croatian MAGIC Consortium, Rudjer Boskovic Institute, University of Rijeka and University of Split, HR-10000 Zagreb, Croatia
\and Max-Planck-Institut f\"ur Physik, D-80805 M\"unchen, Germany
\and Universidad Complutense, E-28040 Madrid, Spain
\and Inst. de Astrof\'isica de Canarias, E-38200 La Laguna, Tenerife, Spain
\and University of \L\'od\'z, PL-90236 Lodz, Poland
\and Deutsches Elektronen-Synchrotron (DESY), D-15738 Zeuthen, Germany
\and ETH Zurich, CH-8093 Zurich, Switzerland
\and Universit\"at W\"urzburg, D-97074 W\"urzburg, Germany
\and Centro de Investigaciones Energ\'eticas, Medioambientales y Tecnol\'ogicas, E-28040 Madrid, Spain
\and Technische Universit\"at Dortmund, D-44221 Dortmund, Germany
\and Inst. de Astrof\'isica de Andaluc\'ia (CSIC), E-18080 Granada, Spain
\and Universit\`a di Padova and INFN, I-35131 Padova, Italy
\and Universit\`a dell'Insubria, Como, I-22100 Como, Italy
\and Unitat de F\'isica de les Radiacions, Departament de F\'isica, and CERES-IEEC, Universitat Aut\`onoma de Barcelona, E-08193 Bellaterra, Spain
\and Institut de Ci\`encies de l'Espai (IEEC-CSIC), E-08193 Bellaterra, Spain
\and Japanese MAGIC Consortium, Division of Physics and Astronomy, Kyoto University, Japan
\and Finnish MAGIC Consortium, Tuorla Observatory, University of Turku and Department of Physics, University of Oulu, Finland
\and Inst. for Nucl. Research and Nucl. Energy, BG-1784 Sofia, Bulgaria
\and Universitat de Barcelona, ICC, IEEC-UB, E-08028 Barcelona, Spain
\and Universit\`a di Pisa, and INFN Pisa, I-56126 Pisa, Italy
\and now at: NASA Goddard Space Flight Center, Greenbelt, MD 20771, USA and Department of Physics and Department of Astronomy, University of Maryland, College Park, MD 20742, USA
% ...
}



   \date{}

  \abstract
   {}
   {Amongst more than fifty blazars detected in very high energy (VHE, E$>$100\,GeV) $\gamma$ rays, only three belong to the subclass of flat spectrum radio quasars (FSRQs). The detection of FSRQs in the VHE range is challenging, mainly because of their soft spectra in the GeV-TeV regime. MAGIC observed PKS~1510$-$089 (z=0.36) starting 2012 February 3 until April 3 during a high activity state in the high energy (HE, E$>$100 MeV) $\gamma$-ray band observed by AGILE and \textit{Fermi}. MAGIC observations result in the detection of a source with significance of 6.0 standard deviations ($\sigma$). We study the multi-frequency behaviour of the source at the epoch of MAGIC observation, collecting quasi-simultaneous data at radio and optical (GASP-WEBT and F-Gamma collaborations, REM, Steward, Perkins, Liverpool, OVRO, and VLBA telescopes), X-ray ({\it Swift} satellite), and HE $\gamma$-ray frequencies.}
   {We study the VHE $\gamma$-ray emission, together with
the multi-frequency light curves, 43\,GHz radio maps, and spectral energy distribution (SED) of the source. 
The quasi-simultaneous multi-frequency SED from the millimetre radio band to VHE $\gamma$ rays is modelled with a one-zone inverse Compton model. We study two different origins of the seed photons for the inverse Compton scattering, namely the infrared torus and a slow sheath surrounding the jet around the Very Long Baseline Array (VLBA) core.}
   {We find that the VHE $\gamma$-ray emission detected from PKS~1510$-$089 in 2012 February-April agrees with the previous VHE observations of the source from 2009 March-April. We find no statistically significant variability during the MAGIC observations on daily, weekly, or monthly time scales, while the other two known VHE FSRQs (3C~279 and PKS~1222+216) have shown daily scale to sub-hour variability. The $\gamma$-ray SED combining AGILE, \textit{Fermi} and MAGIC data joins smoothly and shows no hint of a break. The multi-frequency light curves suggest a common origin for the millimetre radio and HE $\gamma$-ray emission, and the HE $\gamma$-ray flaring starts when the new component is ejected from the 43\,GHz VLBA core and the studied SED models fit the data well. However, the fast HE $\gamma$-ray variability requires that within the modelled large emitting region, more compact regions must exist. We suggest that these observed signatures would be most naturally explained by a turbulent plasma flowing at a relativistic speed down the jet and crossing a standing conical shock.}
   {}

\titlerunning{PKS~1510$-$089}
   \maketitle
%
\section{Introduction}

The most numerous extragalactic very high energy (VHE, E$>$100\,GeV) $\gamma$-ray sources
are blazars, which are active galactic nuclei (AGN) with a relativistic
jets pointing close to our line of sight. Within the blazar group the
most numerous VHE $\gamma$-ray emitters are X-ray bright BL Lacertae objects
(BL Lacs) while only three blazars of the flat spectrum radio quasars (FSRQs)
type have been detected to emit VHE $\gamma$ rays.

Blazars typically show variable emission in all wavebands from radio to
$\gamma$ rays.
FSRQs are more luminous than BL Lacs at $\gamma$ rays
and so they could, in principle, be observed at greater distances at very high energies. 
The SEDs of both types of sources show two
peaks: the first is generally attributed to synchrotron emission and the
second one to inverse Compton (IC) scattering, though hadronic
 mechanisms have also been proposed for producing the second peak
\citep[see e.g.][]{bottcher}. In FSRQs the first peak is usually in the infrared
regime, while for BL~Lacs it is between infrared and hard X-rays. The
optical spectra of FSRQs show broad emission lines, indicative of high velocity gas in the so-called broad line region (BLR)
close (0.1 to 1 parsec) to the
central engine \citep[e.g.][]{kaspi}, while BL Lacs show very weak or no emission
lines in their spectra. Because of these properties FSRQs were not thought to be good candidates
to emit VHE $\gamma$ rays: the
low synchrotron peak frequency may imply efficient synchrotron cooling, which makes
it difficult to produce VHE $\gamma$-ray emission. Additionally, if the $\gamma$ rays are produced close to the central engine, the BLR clouds absorb the $\gamma$-ray emission via pair production. 
The high redshift also implies strong absorption of VHE $\gamma$ rays
by the extragalactic background light \citep[EBL;][]{Ste92,Hau01}. 
Despite these difficulties,
MAGIC detected VHE $\gamma$ rays from the FSRQ 3C~279 (z=0.536) in 2006
\citep{Science}. This discovery was followed by a second
detection in 2007 \citep{3C279} and the detection of two other FSRQs
\object{PKS~1510$-$089} (z=0.36) by H.E.S.S. \citep{hess} in 2009 and PKS~1222+216 (z=0.432) in 2010 \citep{1222}. In this paper we report the detection of VHE
$\gamma$ rays from \object{PKS~1510$-$089} in 2012 February-April \citep{atel}
by the MAGIC telescopes.

The standard picture for FSRQs is that the
$\gamma$ rays are
emitted close to the central black hole (so called ``near-dissipation zone"), where the external photons
from BLR can serve as seed photons for IC scattering
\citep[e.g.][]{hartman}. This picture was already challenged in the
EGRET era by the observations of a connection between radio outburst and
$\gamma$-ray flares \citep[e.g.][]{jorstad01,lahteenmaki,lindfors06}.
The observations of VHE
$\gamma$ rays from FSRQs have further challenged the ``near-dissipation zone''
emission scenario \citep[see e.g.][]{3C279,1222}, because in order to produce the observed VHE $\gamma$-ray flux, the MeV $\gamma$-ray flux would have to be much higher than observed.
Moreover, the combined HE to VHE $\gamma$-ray spectrum does not show a break at
a few tens of GeV 
as would be expected if the $\gamma$ rays originated in inside the BLR \citep[e.g.][]{FM}. In addition, at least in some cases (3C~279 in 2007 and PKS~1222+216 in 2010), the VHE $\gamma$-ray detections were coincident with zero-separation epochs of new knots emerging from the 43\,GHz Very Long Baseline Array (VLBA) core \citep{larionov,jorstad11, marscher12}, suggesting that VHE $\gamma$ rays could be emitted in these knots, tens of parsecs away from the central engine. Arguments for and against the ``near-dissipation zone'' are systematically discussed in e.g. \citet{sikora09}. In general, the main argument against emission originating far away from the central engine has been the fast variability observed in $\gamma$ rays. However, the recent model by \citet{marscher14} where relativistic turbulent plasma crosses a standing shock, could potentially explain both the observed radio-gamma connection and the fast variability of $\gamma$ rays.

\object{PKS~1510$-$089} is a $\gamma$-ray bright quasar, whose jet exhibits one
of the fastest apparent motions (up to 45$c$) amongst all blazars \citep{jorstad05}.
It was discovered in HE $\gamma$ rays by EGRET, but no variability was detected \citep{hartman99}, while in the AGILE and {\it Fermi} era it has shown several flaring epochs.  A variability study of this source with AGILE data in the period
2007 July -- 2009 October was presented in \citet{verrecchia13}.
The source showed bright flares at radio, optical,
X-ray and HE $\gamma$-ray energies at the beginning of 2009 \citep{marscher,abdo10,dammando11}. The discovery of VHE $\gamma$ rays from \object{PKS~1510$-$089} also took place in this period, displaying a rather low flux (F\,($>$150 GeV)=($1.0\pm0.2_{\mathrm stat}\pm0.2_{\mathrm sys})\cdot10^{-11}$ ph cm$^{-2}$ s$^{-1}$, $\sim$3\% of Crab Nebula flux) and a very soft spectrum \citep[with photon index, $\Gamma = 5.4\pm 0.7_{\rm stat}\pm 0.3_{\rm sys}$,][]{hess}. 
In HE $\gamma$ rays this outburst consisted of several flares. In X-rays
flaring was moderate and not correlated with the $\gamma$-ray flaring, but the last $\gamma$-ray flare was accompanied by a large optical outburst (reaching a peak flux of 18\,mJy in the R-band
while the quiescent level flux is typically $\sim2$\,mJy) and a large 
radio outburst (reaching a maximum of 4\,Jy, 1-2\,Jy being the normal quiescent state flux at 37\,GHz). During the $\gamma$-ray flares the optical electric vector
position angle (EVPA) rotated by $>720^\circ$ and during the major
optical flare, the optical polarisation degree increased to $>30\%$. In
the 43\,GHz VLBA maps a superluminal knot emerged from the VLBA core with a
zero-separation epoch essentially simultaneous with this sharp optical
flare. \citet{marscher}
concluded that the $\gamma$-ray flaring activity was taking place in a knot seen in the VLBA images at later times, placing the emission region distant from the central engine. This and the variable synchrotron to $\gamma$-ray ratio require that there are local sources of seed photons for IC scattering within or just outside the jet (e.g. a slow sheath of a jet). 
In contrast, based on the ratio between optical and $\gamma$-ray variability \citet{abdo10} 
concluded that the $\gamma$-ray emission favors an external Compton model where the seed photons are provided by the BLR clouds.    

In second half of 2011 the source again showed activity in
several bands. First, in 2011 July, there were optical and HE $\gamma$-ray
flares accompanied by the rotation of the EVPA by $>380^\circ$ \citep{orienti}. In second half of 2011 \object{PKS~1510$-$089}
underwent the brightest radio flare ever observed from the
source and there was associated high activity in the HE $\gamma$-ray band. The
flare was accompanied by the appearance of a new component in the VLBA
jet at 15\,GHz \citep[][]{orienti} and by extremely fast $\gamma$-ray
variability with time scales down to 20 minutes
\citep[e.g.][]{saito,2013arXiv1304.2878F}. Unfortunately, during this
period the source was not observable for ground based optical and $\gamma$-ray instruments.

In 2012 February \object{PKS~1510$-$089} showed again high activity in HE
$\gamma$ rays \citep{lucarelli}. This triggered observations of
the source with the MAGIC telescopes 
which resulted with a significant detection of VHE $\gamma$ rays \citep{atel, decaneva}. The results from the MAGIC observations
(Section 2) are presented together with HE $\gamma$-ray data from
AGILE and {\it Fermi} (Section 3), X-ray data from {\it Swift} (Section 4), near
infrared, optical, ultraviolet (Section 5), and radio observations
(Section 6) from several instruments. A subset of the data presented
here have been previously presented in \citet{fermisymp}, while in
this paper we present the full analysis of the multi-frequency
behaviour of the source during 2012 February-April and compare it with the
previous flaring epochs of \object{PKS~1510$-$089}.
   
\section{MAGIC VHE $\gamma$-ray observations, data analysis, and results}

\subsection{Observations and data analysis}
MAGIC is a system of two 17\,m diameter Imaging Air Cherenkov
Telescopes (IACTs) located at the Roque de los Muchachos Observatory
on La Palma, one of the Canary Islands (28$^\circ$46$^\prime$ N,
17$^\circ$53.4$^\prime$ W at 2231 m a.s.l.). The large collection area
of the telescopes and the advanced observational techniques enables us to reach a low energy threshold of 50\,GeV (in a normal stereo
trigger mode) at low zenith angles. In late 2011 the telescope readout system was upgraded and replaced
\citep{2013arXiv1305.1007S}.

The MAGIC target of opportunity (ToO) observations of \object{PKS~1510$-$089} were
carried out from 2012 February 3 to April 3 (MJD~55960-56020). During 28 nights
$\sim$25\,hours of data were taken with the stereo trigger, of which
21.4\,hours data passed quality selection.  The data were collected at
zenith angles between 37$^\circ$ and 49$^\circ$.  The telescopes were
operated with the false source tracking method \citep{Fomin94}, the
so-called wobble mode, in which the pointing direction counter-changes
every 20 minutes between four sky positions at 0.4$^\circ$ offset with
respect to the source position. Four wobble positions improve the background statistics, since three OFF positions can be sampled which reduces the impact of inhomogeneities in the camera acceptance.

We analysed the data 
in the MARS analysis framework \citep{Moralejo09}. 
The images were processed using a cleaning algorithm that accounts for timing information \citep{Aliu09}.
The criteria for {\it core} and {\it boundary} 
pixels are eight and four photo-electrons, respectively. These are 
different from those used for the standard analyses done before the upgrade of the 
readout \citep[the details are described in ][]{Performance} mainly due to the
different noise level of the new readout system.
The random forest (RF) method was used for the gamma-hadron separation 
\citep{Albert08_RF} using both mono and stereoscopic parameters. The reconstructed 
shower arrival direction of each telescope was calculated with the RF DISP method 
\citep{Aleksic10}, and the weighted mean of the closest pair amongst the 
reconstructed DISP positions is regarded as the final reconstructed position. 

\subsection{Results}

\begin{figure}
\resizebox{\hsize}{!}{\includegraphics{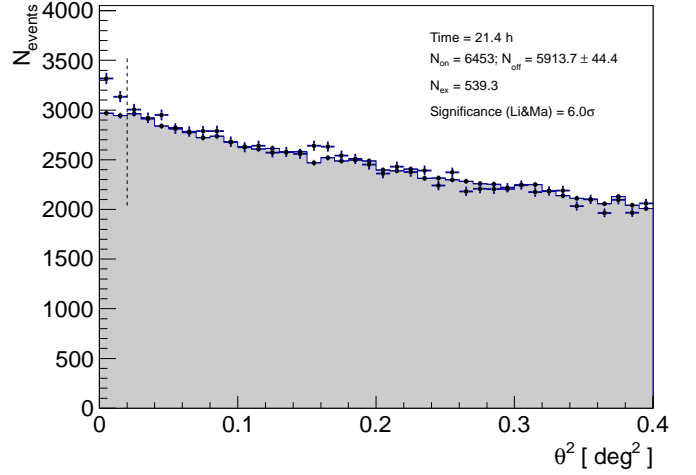}}
 \caption{Distribution of the square of reconstructed shower direction ($\theta^2$) with 
  respect to the position of \object{PKS~1510$-$089} for the ON (black points) and 
  the OFF (grey shaded area) in the camera coordinates. The events inside the vertical dashed line, 
  corresponding to the a priori-defined signal region, are used to compute the detection significance.}
 \label{MAGICfig1}
\end{figure}

The distributions of squared angular distances between the
reconstructed source position and the nominal source position in the
camera, the so-called $\theta^2$ plot, is shown in
Fig.\ref{MAGICfig1}. The number of the background events was extracted
from the three OFF regions which were symmetrical relative to the
pointing position.  Above the normalised background events, an excess
of 539 $\gamma$ rays was found.  The significance of a signal
detection was evaluated following Equation (17) of \citet{Li83}. We
found a corresponding significance of 6\,$\sigma$ from the 21.4\,hours
observational data. The observation at high zenith angle had a
somewhat higher energy threshold of 120\,GeV, determined from the
Monte Carlo rate with an assumed photon index of 4.0.

To derive the energy spectrum of \object{PKS~1510$-$089} the unfolding procedure
\citep{Albert07} was performed to correct for a distortion introduced by the 
detector which has a finite resolution and biases. Moreover, absorption by 
e$^+$e$^-$ pair creation due to the interaction with the EBL photons 
was also corrected for through the same unfolding process, using one of the several state-of-the-art  EBL model \citep{dominguez}. 

\begin{figure}
 \resizebox{\hsize}{!}{\includegraphics{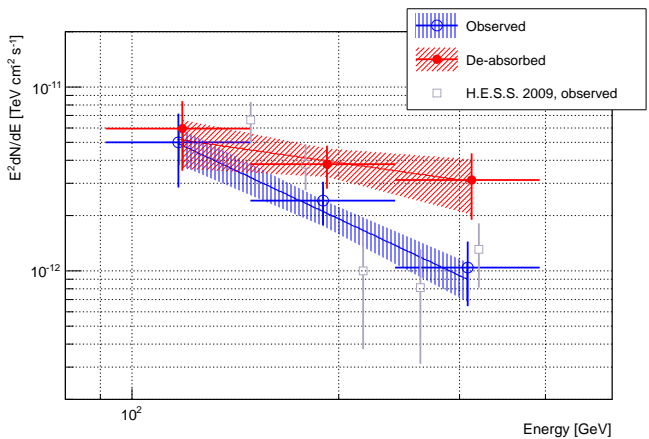}}
 \caption{VHE differential energy spectra of \object{PKS~1510$-$089} measured by MAGIC in the period between 2012 February 3 and April 3. The blue open circles and the blue shaded region show the observed spectrum and its statistical uncertainty, the red dots and the red shaded region show the de-absorbed spectrum (see text). The grey open squares are the source spectrum observed in March-April 2009 by H.E.S.S. \citep{hess}.}
 \label{MAGICfig2}
\end{figure}

We found that different unfolding methods gave consistent results, and the energy
spectrum before the EBL correction can be well reproduced by a power law
\begin{equation}
 \frac{dF}{dE} = F_0\left(\frac{E}{200\,\mathrm{GeV}}\right)^{-\Gamma},
\end{equation}
where $F_0 =(4.8\pm0.9_{\rm stat}\pm 1.3_{\rm sys})\, \times 10^{-11}\,\mathrm{cm^{-2}\,s^{-1}\,TeV^{-1}}$ 
and $\Gamma = 3.8 \pm 0.4_{\rm stat} \pm 0.3_{\rm sys}$ are the flux constant at 
200\,GeV and the photon index, respectively.
As \object{PKS~1510$-$089} is a very weak, steep spectrum VHE source the systematic errors are larger than the ones evaluated in \citet{Performance}.  
The systematic error in the energy scale is 17\,\% as in \citet{Performance}.
Fig.\ref{MAGICfig2} shows the differential energy spectra of \object{PKS~1510$-$089} measured by MAGIC in 2012. 
The fitted function and its one sigma error range displayed as the shaded regions were obtained through 
the forward unfolding, and the spectral points were derived using the Bertero unfolding method \citep{Bertero}.
The spectrum extends up to $\sim$400\,GeV. 
The integral flux above 120\,GeV was estimated to be 4\,\% of the Crab Nebula's flux. After 
the correction for the EBL attenuation the spectrum is still well fitted by a 
power law with an intrinsic photon index of $\Gamma_{\rm int} = 2.5\pm 0.6_{\rm stat}$. 
The flux and spectrum are in agreement with those observed by H.E.S.S. in March-April 2009 \citep{hess}.

The $\gamma$-ray flux variability above 200\,GeV was studied on both daily and weekly 
time scales. The mean flux above 200\,GeV of \object{PKS~1510$-$089} in this period was
F\,($>$200 GeV)=(3.6$\pm$0.9)$\times10^{-12}$ ph cm$^{-2}$ s$^{-1}$.
The reduced $\chi^2$ of the fit with a constant flux is $\chi^2/n_{dof}=40.5/24$ (2.3\,$\sigma$) for daily and $\chi^2/n_{dof}=7.7/4$ (1.6\,$\sigma$) for weekly light curve, consistent with no statistically significant variability. 
Following the method used in \citet{2344} we also estimated
how much variability could be hidden in the data. We derived
a 3\,$\sigma$ confidence level upper limit for individual nights$/$weeks and compared it to the observed mean flux adopting the night-to-night systematic error of 12\% \citep{Performance}. We found that variability of a factor of eight in nightly scale and factor of 2.5 in the weekly scale could be missed. 
The weekly light curve is displayed and discussed 
with the multi-frequency data in Section 7.
The observed VHE $\gamma$-ray emission, showing only marginal variability over several weeks, displays a different behaviour than other FSRQs
\citep{Science,1222,3C279}, but is in agreement with previous observations of \object{PKS~1510$-$089} by H.E.S.S. in 2009 March-April \citep{hess}.

\section{HE $\gamma$-ray observations, data analysis, and results}

We investigate the emission in the HE $\gamma$-ray range making use of two instruments: \agile{}-GRID and {\it Fermi}--LAT. The comparison and combination of the HE and VHE $\gamma$-ray results are presented in Section 3.3. 

\begin{table*}
\centering
\caption{Integral photon fluxes $>$100 MeV detected by AGILE-GRID}
\begin{tabular}{ccccc}
\hline
\hline
\multicolumn{1}{c}{Epoch} &
\multicolumn{1}{c}{Integration period} &
\multicolumn{1}{c}{Energy bin} &
\multicolumn{1}{c}{Flux}&
\multicolumn{1}{c}{$\Gamma$} \\
 \multicolumn{1}{c}{ } &
 \multicolumn{1}{c}{[MJD]} &
 \multicolumn{1}{c}{[MeV]} &
 \multicolumn{1}{c}{$[\rm{ph~cm^{-2} s^{-1}}]$ } &
 \multicolumn{1}{c}{ }\\
\hline
Flare-I   (7 days)  & 55952.5 - 55959.5 & $>$ 100 & $(2.0\pm0.5)\times 10^{-6}$ & $2.17\pm0.24$\\
Flare-II  (14 days) & 55977.5 - 55991.5 & $>$ 100 & $(4.4\pm0.5)\times 10^{-6}$ & $2.21\pm0.11$\\
Postflare (14 days) & 55998.5 - 56012.5 & $>$ 100 & $(1.8\pm0.5)\times 10^{-6}$ & $2.39\pm0.36$\\
Low$/$intermediate state & 55746.5 - 55803.5 & $>$ 100 & $(9.1\pm1.5)\times 10^{-7}$ & $2.44\pm0.17$\\
\hline
\end{tabular}
\label{tab1:AGILEresults}
\end{table*}

\begin{table*}
\centering
\caption{Differential flux values ($\rm{\nu F(\nu)}$) detected by AGILE-GRID in 2012}
\begin{tabular}{ccccc}
\hline
\hline
\multicolumn{1}{c}{Epoch} &
\multicolumn{1}{c}{Integration period} &
\multicolumn{1}{c}{Energy bin} &
\multicolumn{1}{c}{ $\nu$ } &
\multicolumn{1}{c}{$\rm{\nu F(\nu)}$}\\
 \multicolumn{1}{c}{ } &
 \multicolumn{1}{c}{[MJD]} &
 \multicolumn{1}{c}{[MeV]} &
 \multicolumn{1}{c}{[Hz]} &
\multicolumn{1}{c}{$[\rm{erg~cm^{-2} s^{-1}}]$}\\
\hline
Flare-II  (14 days) & 55977.5 - 55991.5 & 100 - 200    & 3.42 $\times 10^{22}$ & $(7.0\pm1.1)\times 10^{-10}$ \\
                    &                   & 200 - 400    & 6.85 $\times 10^{22}$ & $(7.2\pm1.3)\times 10^{-10}$ \\
                    &                   & 400 - 10000  & 4.84 $\times 10^{23}$ & $(5.2\pm1.1)\times 10^{-10}$ \\
\hline
Low$/$intermediate state & 55746.5 - 55803.5 & 100 - 200    & 3.42 $\times 10^{22}$ & $(1.7\pm0.4)\times 10^{-10}$ \\
                    &                            & 200 - 400    & 6.85 $\times 10^{22}$ & $(1.7\pm0.4)\times 10^{-10}$ \\
                    &                            & 400 - 10000  & 4.84 $\times 10^{23}$ & $(6.4\pm0.3)\times 10^{-11}$ \\
\hline
\end{tabular}
\label{tab2:AGILEresults}
\end{table*}

\subsection{AGILE}

AGILE \citep[Astrorivelatore Gamma ad Immagini LEggero,][]{Tavani2009:AGILE} 
is a scientific mission 
devoted to the observation of astrophysical sources of
HE $\gamma$ rays in the 30\,MeV -- 30\,GeV energy range,
with simultaneous X-ray imaging capability in the 18 -- 60\,keV
band. The \agile{} payload combines for the first time two coaxial detectors:
the gamma-ray imaging detector (GRID, composed of
a 12-plane silicon-tungsten tracker, a cesium-iodide mini-calorimeter
and an anti-coincidence shield) and the hard X-ray
detector Super-AGILE. 
The $\gamma$-ray GRID imager provides good performance in a relatively small and compact instrument due to the use of silicon technology:
an effective area of the order of 500 cm$^2$ at several hundred MeV, an
angular resolution of around 3.5$^\circ$ at 100\,MeV, decreasing below
1$^\circ$ above 1\,GeV, a very large field of view ($\sim$ 2.5 sr), as well as
accurate timing, positional and altitude information.

During the first $\sim 2.5$ years 
(2007 July - 2009 October), 
AGILE was operated in
``pointing observing mode'', characterised by long observations
called observation blocks (OBs), 
typically of two to four weeks duration.
Since 2009 November 4, following a malfunction of the rotation 
wheel, AGILE is operating in  ``spinning observing mode'', 
surveying a large fraction (about $70 \% $) of the sky each day.
Thanks to its sky monitoring capability and fast ground segment alert system
distributed amongst the AGILE Data Centre (ADC) and the AGILE team institutes,
AGILE is very effective in detecting bright $\gamma$-ray flares
from blazars. 

Data were analysed applying the
AGILE maximum likelihood (ML) 
analysis on the \object{PKS~1510$-$089} sky position, using the 
standard level-3 AGILE-GRID archive at ADC. 
This archive is composed by 
counts, exposure and diffuse $\gamma$-ray background \citep{Giuliani2004:diff_model} maps generated on several time scales (one day, one week, 28 days)
from the official level-2 data archives, publicly available at the
ADC site\footnote{
ADC pointing (sw=5\_19\_18\_17) and spinning (sw=5\_21\_18\_19) archives, from
\texttt{http://agile.asdc.asi.it}}.
Maps were generated for E $>$ 100~MeV 
including all events collected up to $60^{\circ}$ off-axis,
excluding south Atlantic anomaly data, and
by excluding regions within $10^{\circ}$ from the Earth limb to reduce albedo contamination.
The data have been processed with the latest available software and 
calibrations\footnote{AGILE\_SW\_5.0\_SourceCode from ADC website, 
with \texttt{I0023} calibrations.}.
For a general description of the \agile{} data reduction and of the standard analysis pipeline see \citet[]{Pittori2009:Catalogue,vercellone}. Systematic errors
of the AGILE ML analysis have been estimated to be $\sim$10\% of the flux values \citep{bulgarelli}.

\par
At the beginning of 2012, AGILE detected the \object{PKS~1510$-$089} 
in a high state in two distinct periods: 
one at the end of January-beginning of February, and the other
at the end of February-beginning of March.
The AGILE-GRID (E$>$100\,MeV) light curves covering the MAGIC 
observation of \object{PKS~1510$-$089} from January to March (MJD~55960-56000), with
two days time binning 
are shown together with the multi-frequency light curves in Section 7.
In comparing the AGILE and {\it Fermi} light curves it should be taken
into account that over short time intervals,
AGILE might not spectrally resolve the blazar due to low statistics, and in such cases a ``standard'' fixed
spectral photon index value of 2.1 is adopted for the ML analysis.
This effect may result in an additional systematic error
on the flux (not shown in the figure).
By using, for example, a fixed spectral index value of 2.4, AGILE flux values
would change on average by a factor +15\%.


The first high state (Flare-I) triggered 
the AGILE alert system and four day quick-look results were reported in
ATel \#3907~\citep{ATel3907}. 
Performing a refined ML analysis by optimizing the background estimates
on the AGILE-GRID data covering the seven day period from January 26 to February 2 (MJD~55952.5 to 55959.5),
yields in a detection at a significance level of about 7\,$\sigma$.
The flare-I spectral analysis gives a photon index 
$\Gamma=2.17 \pm 0.24$ 
and a flux 
F\,(E $>$ 100 MeV)=$(2.0 \pm 0.5)\cdot10^{-6}$~ph~cm$^{-2}$~s$^{-1}$.

The second flare (flare-II), with higher $\gamma$-ray flux, was
announced with ATel \#3934~\citep{lucarelli}. The source
maintained its high state above
4.0$\cdot$10$^{-6}$~ph~cm$^{-2}$~s$^{-1}$ for almost two weeks.
We performed the AGILE ML analysis on this
two-week period (from 2012 February 20 to 2012 March 05, MJD~55977.5
to 55991.5) obtaining a detection at a $\sim$16\,$\sigma$ significance
level. The corresponding spectral analysis provides a photon index
$\Gamma$=2.21$\pm$0.11, consistent with that of Flare-I, but a higher flux
F\,(E$>$100~MeV)=(4.4$\pm$0.5)$\cdot$10$^{-6}$~ph~cm$^{-2}$~s$^{-1}$.

After 2012 March 9 (MJD~55995) the source went back to a low-flux state,
with the source sky position approaching the 
border of field of view of AGILE, and after 2012 March 14 (MJD~56000)
the AGILE daily effective exposure gradually decreased.
The ML analysis over 
the 14 day period starting on 2012 March 12 (MJD~55998.5) gives the source at a 
significance level of around 6\,$\sigma$, with a 
photon index $\Gamma$=2.4$\pm$0.4 and an average flux 
F\,(E$>$100~MeV)=(1.8$\pm$0.5)$\cdot$10$^{-6}$~ph~cm$^{-2}$~s$^{-1}$. 

For comparison, 
we have identified one of the 
typical low$/$intermediate states of the source 
with $\gamma$-ray flux below 10$^{-6}$~ph~cm$^{-2}$~s$^{-1}$,
from 2011 July 4 to 2011 August 30 (MJD~55746.5 to 55803.5),
and performed the AGILE ML analysis 
getting a photon index $\Gamma$=2.44$\pm$0.17 and 
a flux F\,(E$>$100~MeV)=(0.91$\pm$0.15)$\cdot$10$^{-6}$~ph~cm$^{-2}$~s$^{-1}$.
AGILE results during the MAGIC observation period in 2012 and this low intermediate state are summarised in Table~\ref{tab1:AGILEresults} and Table~\ref{tab2:AGILEresults} .

\subsection{Fermi-LAT}

\begin{table*}[t]
\centering
\caption{Comparison of the different spectral models for the Fermi-LAT data for \object{PKS~1510$-$089}}
\scalebox{0.85}{
\hspace{-1.5cm}
\begin{tabular}{cccccccccccc@{}c@{}}
\hline\hline
\begin{tabular}{c} \\ \large \bf Epoch \\ \\  \end{tabular}  &
\multicolumn{5}{c}{\large \bf Power-law}   & \multicolumn{6}{c}{\large \bf Log parabola} & \multicolumn{1}{l}{\large \bf $\sigma$\tablefootmark{b}} \\
\cline{2-5} \cline{7-11}
  &         Flux\tablefootmark{a}& 								Index & 					TS & Loglike  && Flux\tablefootmark{a} & Alpha & Beta & TS & Loglike & \\

 \begin{tabular}{c} MAGIC \\ observation \end{tabular} & 3.97 $\pm$ 0.08&2.39 $ \pm$ 0.02 & 12241 &107077 &&3.82 $\pm $ 0.08&2.24 $\pm$ 0.03& 0.09 $\pm$ 0.02 &12243&107056 &6 \\
\vspace{3pt}
 Mean state &  2.67 $\pm$ 0.04 &2.40 $\pm$0.01 &  19943 &269334 && 2.56$\pm$ 0.04 &2.26$\pm$ 0.02 & 0.09 $\pm$ 0.01  &19942&269299& 8\\
\vspace{3pt}
\begin{tabular}{l} Low state \\  \end{tabular} & 0.79 $\pm$ 0.04 &2.52 $\pm$ 0.04 &1417& 99964&&0.75 $\pm$0.04&2.35 $\pm$0.07 &0.12 $\pm$0.04&1422&99959&  3 \\
\vspace{3pt}
\begin{tabular}{l} High state\\ \end{tabular} &6.50$\pm$0.14&2.29$\pm$0.02& 7389&41207&& 6.24$\pm$0.17&2.12$\pm$0.04&0.10$\pm$0.02&7421&41191&4.5\\ \hline\hline
\end{tabular}}
\tablefoot{
\tablefoottext{a}{Flux (100 MeV - 300 GeV ) is in units of [$10^{-6}$ ph cm$^{-2}$s$^{-1}$]}
\tablefoottext{b}{Significance by which the Log parabola model has to be preferred w.r.t. the simple power-law model ($\sigma$ calculated as  [2\,$(Loglike _{Pwl}- Loglike  _{LogP}) ]^{1/2}$).}
}
\label{1510fit}
\end{table*}

\textit{Fermi}--LAT (Large Area Telescope) is a pair conversion telescope designed to cover the energy band from 20\,MeV to greater than 300\,GeV \citep{atwood09}. In its primary observation strategy, survey mode, the LAT scans the entire sky every three hours and therefore can provide observations of \object{PKS~1510$-$089} simultaneous to MAGIC.

PKS~1510$-$089 has been continuously monitored by \textit{Fermi}  and the data  used for this analysis were collected from 2012 January 1 to April 7 (MJD~55927-56024). They were analysed with the standard analysis tool {\it gtlike}, part  of the \textit{Fermi} ScienceTools software package (version 09-27-01). 
Only good quality events within $10^{\circ}$ of PKS~1510$-$089 were selected for analysis. Moreover, to reduce the contamination from the Earth-limb $\gamma$ rays produced by cosmic ray interactions with the upper atmosphere, data were restricted to a maximal zenith angle of $100^{\circ}$ and time periods when the spacecraft rocking angle exceeded $52^{\circ}$ were excluded.

To extract the spectral information we used the standard background models provided by the publicly available files gal\_2yearp7v6\_v0\_trim.fits and    iso\_p7v6source.txt\footnote{\texttt{http://fermi.gsfc.nasa.gov/ssc/data/access/lat/\\BackgroundModels.html}}.
The background templates, whose normalizations were left free during the fitting process, take into account the diffuse $\gamma$-ray emission from our Galaxy and an isotropic diffuse component.
During the spectral fitting of the point source the normalizations of the components comprising the entire background model were allowed to vary freely. 
To derive the source spectral information an unbinned maximum likelihood technique was applied to events in the energy range from 100\,MeV to 300\,GeV \citep{mattox96} in combination with the post-launch instrument response functions 
P7SOURCE\_V6. 
Sources from the 2FGL catalogue \citep{nolan} located within $15^{\circ}$ of PKS 1510$-$089 were incorporated in the model of the region by setting the spectral models and the initial parameters for the modelling to those reported in the 2FGL catalogue. 
In particular, the source of interest was modelled with a Log parabola spectrum \footnote{\texttt{http://fermi.gsfc.nasa.gov/ssc/data/analysis/\\scitools/source\_models.html}}:
\newcommand{\di}{\mathrm{d}}

\begin{equation}
\frac{\di N} {\di E}\,=\,N_0 \left( \frac{E}{E_b} \right) ^{- \left( \alpha+\beta {\mathrm log} (\frac{E}{E_b}) \right)} 
\end{equation}
 where $N_0$ is the normalization, $E_b$ the break energy and $\alpha$ and $\beta$ parameters for the log parabola fit.


In the fitting procedure the parameters of sources located within a $10^{\circ}$ radius centred on the source of interest were left free to vary 
while parameters of sources located within a $10^{\circ}$-$15^{\circ}$ annulus were fixed. When performing the fit for the light curve and SED bins,
 the photon indices of the sources were frozen to the best-fit values obtained from a long-term analysis.
Systematic uncertainties in LAT results due to uncertainties in the effective area are discussed in \citet{ackermann12}; they are smaller than the statistical uncertainties of the points in the light curves and have been neglected.

The \textit{Fermi}-LAT one day bin light curve is shown together with the multi-frequency light curves in Section 7. 
Since the source is not always significantly resolved, flux upper limits at 95\% confidence level were calculated for each time bin where the test statistic{\footnote{a maximum likelihood test statistic TS = 2$\Delta$log(likelihood) between models with and without a
point source at the position of PKS 1510--089 \citep{mattox96}}} (TS) value for the source was TS$<$25.
The light curve shows that the flaring activity had a duration of about 55 days in $\gamma$ rays and consisted of several distinct flares. 

As \object{PKS~1510$-$089} is known to show variability on time scale less than a day \citep{saito,brown} we also searched for shorter time scale of variability within the brightest flaring epoch 2012 February 17 to March 8 (MJD~55974-55994) and produced light curves in bins of 1.5 hours and 3 hours (the latter is shown in Fig.~\ref{combined_lightcurve}).
We systematically looked at the light curves and calculated
the doubling times ($t_d$) between significant (TS$>$25) adjacent bins
following 
$t_d = \Delta t\cdot~\ln2 / \ln(F_{max}/F_{min})$.
Excluding flux variations that were within 1$\sigma$ 
and doubling times with errors larger than 50\%, 
the shortest value that we derive for this period is 
$t_d = 1.5\pm0.6$ hour.

We considered the \emph{Fermi}-LAT data of individual light-curve
bins, fitting them with a power-law model in order to investigate
spectral evolution in the HE range. 
In this analysis we do not find evidence for this behaviour, although we
note that the source spectrum is better represented by the log parabola
shape in several time intervals, thus the power-law fit may not
adequately reproduce the source spectral shape. Additionally it is
apparent that during the high state, the spectral index is
significantly harder than for the low state or mean state (see below).

The SED was obtained combining all events of time intervals coincident
with the last two VHE detections, i.e. from February 19 to March 5
(MJD~55976-55991) and from March 15 to April 3 (MJD~56001-56020). For
comparison we analysed the mean state in 2012 January-April
(MJD~55927-56025), a low state SED which consists of the data taken
in 2012 January and April (MJD~55927-55954 and 56007-56025) and a high
state which consists of all time periods when the {\it Fermi} flux was
$>6\cdot10^{-6}$~ph~cm$^{-2}$ s$^{-1}$.
The log parabola model is significantly preferred (in the MAGIC observing epoch with 6$\sigma$ significance and in the low state with 3$\sigma$) with respect to the power-law
in all the time intervals considered for this SED analysis.
The detailed results are shown in Table~\ref{1510fit}. 

\subsection{Gamma-ray results}

\begin{figure}
\includegraphics[width=0.18\textwidth, angle=-90]{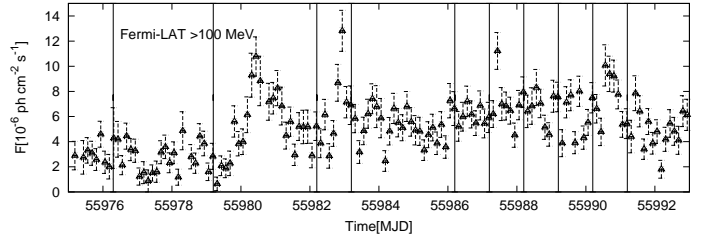}
\caption{{\it Fermi}-LAT $>$100\,MeV light curve in the three hour bins for the first MAGIC observing period. The vertical lines represent the MAGIC observing times (all shorter than three hours in duration) showing that the MAGIC observation windows missed the times of the fastest HE $\gamma$-ray variability.}
\label{combined_lightcurve}
\end{figure}

We compared the results of the observations in HE and VHE $\gamma$ rays. 
As discussed in previous sections, the HE $\gamma$-ray flux is
variable on time scales shorter than day. Therefore it appears that
fast variability can explain the small mismatches between daily fluxes
of {\it Fermi}-LAT and two-day fluxes by AGILE-GRID. These light curves
are shown together with multi-frequency light curves in Section 7. The variability
amplitude of the HE $\gamma$-ray flux is rather large 
(more than one order of
magnitude in flux) in the first MAGIC observing period (MJD~55976 to 55991). Still, within this period, MAGIC observed no statistically significant variability from the source. In Fig.~\ref{combined_lightcurve} the {\it Fermi}-LAT light curve in three hour bins is shown. The vertical lines show the
MAGIC observation times, revealing that the MAGIC observations missed
all the periods of fast HE $\gamma$-ray variability and therefore it
was to be expected that no fast variability would be detected in the MAGIC
observations. Apparently the MAGIC observations also missed the
highest peaks of the HE $\gamma$-ray light curve. The maximum flux
measured simultaneous to the MAGIC observations is F\,($>$100MeV)$\sim8\cdot10^{-6}$ ph~cm$^{-2}$ s$^{-1}$ and the average of the strictly simultaneous bins is F\,($>$100MeV)$\sim$4.4$\cdot10^{-6}$ ph~cm$^{-2}$ s$^{-1}$.

For the second MAGIC observation window in March-April (from 56001 to
56020), fast variability could not be investigated because of
the lower HE $\gamma$-ray state of the source. After March 23 (MJD~56009),
the source was no longer detected on daily scales in HE $\gamma$
rays, the daily upper limits being below $1.0\cdot10^{-6}$ ph~cm$^{-2}$
s$^{-1}$. Therefore, in total, the HE $\gamma$-ray flux variability
amplitude, within the windows strictly simultaneous to the MAGIC observing
windows, was $\sim$eight on nightly scales, which could go
undetected in the MAGIC light curve given the overall low flux as
  discussed in Section 2.2. It is therefore not possible to conclude
  if the lack of significant variability in the VHE $\gamma$-ray band
  has a real physical origin or if it is simply an observational bias (either
  due to unfortunate sampling or due to low photon statistics).

\begin{figure}
\includegraphics[width=0.48\textwidth]{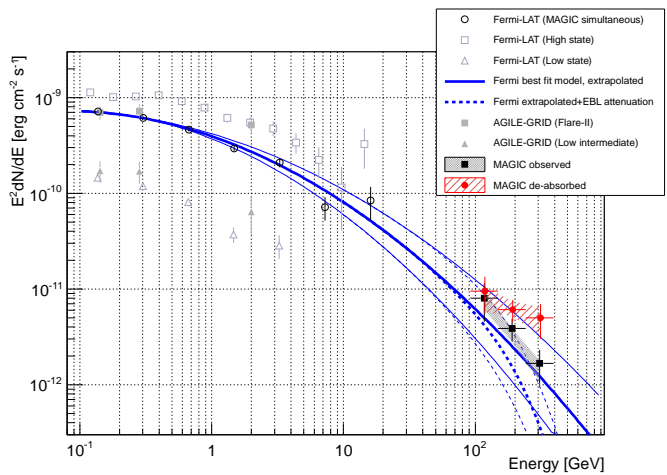}
\caption{$\gamma$-ray SED constructed from AGILE, {\it Fermi}-LAT and MAGIC data. The AGILE-GRID data (grey filled squares) correspond to the data of Flare-II (from MJD~55977.5 to 55991.5). 
The {\it Fermi}-LAT spectrum (black open circles) combines all events of time intervals coincident with the MAGIC observation window (MJD~55976 to 55991 and from 56001 to 56020) 
with the blue lines showing the log parabola fit to the data and its statistical
uncertainty (the thinner lines). The fit and the errors of the {\it Fermi}-LAT spectra have been extrapolated to MAGIC energy range. The dashed blue lines show the extrapolation with the EBL absorption effects included. The MAGIC data points are shown with black filled squares (observed) and red filled circles (de-absorbed). The corresponding shaded region indicates the statistical uncertainty of the spectral fitting (same as in the Fig.~2). The grey data shows, for the comparison, the low-intermediate state spectrum of the source as measured by AGILE-GRID (triangles) and {\it Fermi}-LAT (open triangles) and high-state SED as measured by {\it Fermi}-LAT (open squares).}
\label{SED_Fermi_MAGIC}
\end{figure}

The SED of \object{PKS~1510$-$089} from $\sim$100\,MeV to $\sim$400\,GeV is presented in Fig.~\ref{SED_Fermi_MAGIC}. The HE $\gamma$-ray data from AGILE-GRID and {\it Fermi}-LAT cover slightly different periods (AGILE from MJD~55977.5 to 55991.5 and {\it Fermi}-LAT from MJD~55976 to 55991 and from 56001 to 56020). The AGILE-GRID data consist of flaring state data only while the {\it Fermi}-LAT spectrum summarises all events of the time intervals coincident with the MAGIC observation window. 
As suggested by AGILE and confirmed by {\it Fermi}-LAT, the
brighter states are characterised by a hardening of the HE spectrum, and therefore the higher flux observed by AGILE at 2\,GeV is expected.
The peak of the SED is at $\sim$ 100\,MeV. The log parabola fit and the errors of the {\it Fermi}-LAT spectra have been extrapolated to the MAGIC energy range. We also show the extrapolation taking into account the EBL absorption using the model of \citet{dominguez}. The VHE $\gamma$-ray spectrum observed by MAGIC connects smoothly with this extrapolation suggesting that the emission originates in the same region.     

\section{Swift X-ray observations, data analysis and results}

The {\it Swift} satellite \citep{gehrels04} performed 23 ToO observations on PKS~1510$-$089 between 2012 February 2 and April 5 (MJD~55959-56022), triggered by the strong activity of the source detected first by
AGILE  \citep{lucarelli} and {\it Fermi}-LAT at HE $\gamma$-ray energies, and then by
MAGIC at TeV energies \citep{atel}. The observations were performed with all three onboard instruments: the X-ray Telescope (XRT; \citet{burrows05}, 0.2--10.0 keV), the Ultraviolet Optical Telescope (UVOT; \citet{UVOT}, 170--600 nm), and the Burst Alert Telescope (BAT; \citet{barthelmy05}, 15--150 keV).

For the {\it Swift}-XRT data analysis, we considered observations with
exposure times longer than 500 seconds, including 20 observations. In
addition we summed the data of the three observations performed on
February 19 in order to have higher statistics. The XRT data were
processed with standard procedures (\texttt{xrtpipeline v0.12.6}),
filtering, and screening criteria by using the \texttt{Heasoft}
package (v6.11). The source count rate was low during the entire
campaign ($<$ 0.5 counts s$^{-1}$), so we only considered photon
counting data and further selected XRT event grades
0--12. Pile-up correction was not required. Source events were
extracted from a circular region with a radius of 20 pixels (one pixel
$\sim$ 2.36$"$), while background events were extracted from a 50 pixel
radius circular region not containing any contaminating sources and lying away from the source region. 
The spectral redistribution
matrices v013 in the Calibration database maintained by HEASARC were used.

The adopted energy range for spectral fitting is 0.3--10\,keV. When the number
of counts was less than 200 the Cash statistic \citep{cash79} on ungrouped data
was used. All the other spectra were rebinned with a minimum of 20 counts per energy bin to allow $\chi{^2}$ fitting within
{\sc XSPEC} \citep[v12.6.0;][]{arnaud96}. We fitted the individual spectra with a
simple  absorbed power law, with a neutral hydrogen column density fixed to its
Galactic value \citep[6.89 $\cdot$ 10$^{20}$ cm$ ^{-2}$;][]{kalberla05}. The fit results are reported in Table~\ref{XRT_1510}.

During the observations {\it Swift}-XRT detected the source with a flux, F\,(0.3--10 keV),
 in the range (0.7-1.2)$\cdot$10$^{-12}$ erg cm$^{-2}$ s$^{-1}$, comparable to the flux observed in 2009 March, during a period of high HE $\gamma$-ray activity \citep{dammando11, abdo10}, but lower with respect to the high flux level observed in 2006 August \citep{kataoka08}. The light curve is shown in Section 7, together with the multi-frequency data.

\begin{figure}
\includegraphics[width=0.55\textwidth]{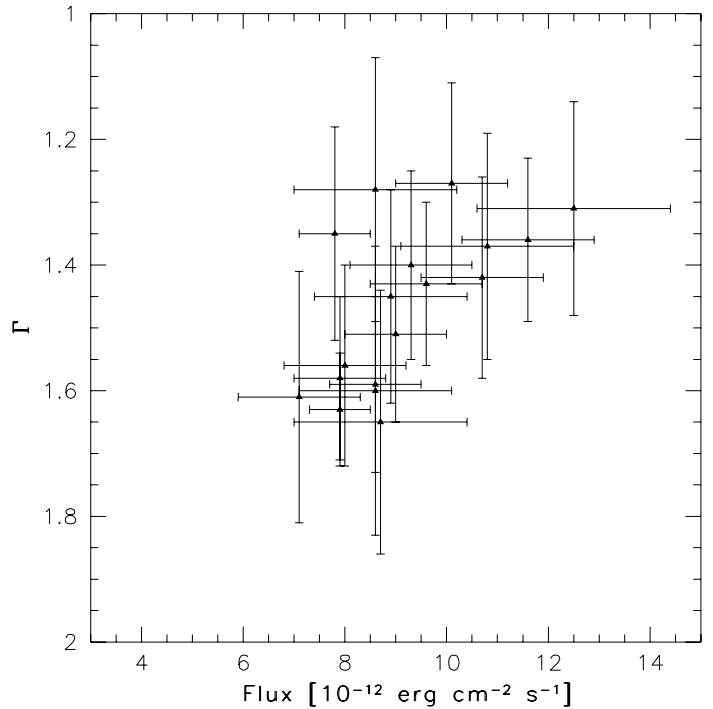}
\caption{Flux (0.3--10 keV) versus photon index for {\it Swift}-XRT. Although there was only marginal X-ray variability during the observations, the plot shows a hint of harder when brighter trend.}
\label{Swift_hardness}
\end{figure}

The flux versus photon index plot is shown in
Fig.~\ref{Swift_hardness}. At higher flux the photon index seems to become harder.
This behaviour is consistent with the harder when brighter trend reported in \citet{kataoka08} and \citet{dammando11}. As discussed in these papers, such a trend indicates that in bright states the X-ray emission is completely dominated by external Compton emission, while in lower state there is also contribution from a soft excess component, which could be e.g. a blurred reflection,
Comptonization of the thermal disc emission or a mixture of synchrotron, external Compton and SSC emission. 

We also investigated the {\it Swift}-BAT data using the {\it Swift}-BAT Hard X-ray Transient Monitor \citep{krimm}. In the BAT
data for 2012 January-April there is only a hint of signal
(2.5$\sigma$) on 2012 February 9 (MJD~55966), with a rate of
($0.0033\pm0.0013$) counts s$^{-1}$ cm$^{-2}$, corresponding to
15\,mCrab in the 15-50\,keV energy band.  As a comparison, in 2009
March the high flux observed by BAT in hard X-ray was 40\,mCrab
\citep{dammando11}.

\begin{table*}[th!]
%
\centering
\caption{Log and fitting results of {\it Swift}-XRT observations}
\begin{tabular}{ccccc}
\hline
\hline
\noalign{\smallskip}
\multicolumn{1}{c}{Date} &
\multicolumn{1}{c}{Net Exp. Time} &
\multicolumn{1}{c}{Photon Index} &
\multicolumn{1}{c}{Flux 0.3--10.0 keV\tablefootmark{a}} &
\multicolumn{1}{c}{$ \chi^{2}_{\rm red}$ (d.o.f.) / Cash} \\
\multicolumn{1}{c}{} &
\multicolumn{1}{c}{ (sec) } &
\multicolumn{1}{c}{$\Gamma$}&
\multicolumn{1}{c}{($\times$ 10$^{-12}$ erg cm$^{-2}$ s$^{-1}$}) &
\multicolumn{1}{c}{} \\
\hline
\noalign{\smallskip}
$2012-02-02$ &  2470   &  $1.35 \pm 0.17$ &	 $7.8 \pm 0.7$   &   Cash   \\
$2012-02-04$ &  2450   &  $1.42 \pm 0.16$ &	$10.7 \pm 1.2$   &   0.85 (19)  \\
$2012-02-05$ &  2655   &  $1.27 \pm 0.16$ &	$10.1 \pm 1.1$   &   1.00 (18)  \\
$2012-02-07$ &  2140   & 	$1.56 \pm 0.16$ & $8.0 \pm 1.2$   &   1.00 (14) \\ 
$2012-02-17$ &  789    &  $1.65 \pm 0.21$ &	 $8.7 \pm 1.7$   &   Cash        \\ 
$2012-02-19$ &  5781   & 	$1.63 \pm 0.09$ &	 $7.9 \pm 0.6$   &   0.95 (39) \\
$2012-02-21$ &  1286   & 	$1.60 \pm 0.23$ &	 $8.6 \pm 1.5$   &   0.76 (8)   \\
$2012-02-22$ &  2700   & 	$1.51 \pm 0.14$ &	 $9.0 \pm 1.0$   &   1.05 (19)  \\
$2012-02-23$ &  2989   & 	$1.43 \pm 0.13$ &	 $9.6 \pm 1.1$   &   0.85 (22) \\
$2012-03-01$ &  1024   & 	$1.37 \pm 0.18$ &	$10.8 \pm 1.7$   &   Cash        \\
$2012-03-18$ &  3224   & 	$1.36 \pm 0.13$ &	$11.6 \pm 1.3$   &   0.77 (20) \\ 
$2012-03-20$ &  1351   & 	$1.45 \pm 0.17$ &	 $8.9 \pm 1.5$   &   Cash        \\
$2012-03-22$ &  2477   & 	$1.28 \pm 0.21$ &	 $8.6 \pm 1.6$   &   1.06 (9)   \\
$2012-03-24$ &  1219   & 	$1.31 \pm 0.17$ &	$12.5 \pm 1.9$   &   Cash        \\
$2012-03-30$ &  2695   & 	$1.58 \pm 0.13$ &	 $7.9 \pm 0.9$   &   1.01 (17)  \\
$2012-04-01$ &  2620   & 	$1.59 \pm 0.14$ &	 $8.6 \pm 0.9$   &   0.71(17)  \\
$2012-04-03$ &  1596   & 	$1.40 \pm 0.15$ &	 $9.3 \pm 1.2$   &   Cash        \\
$2012-04-05$ &  1196   & 	$1.61 \pm 0.20$ &	 $7.1 \pm 1.2$   &   Cash        \\
\noalign{\smallskip}
\hline
\hline
\noalign{\smallskip}
\end{tabular}
\\
\tablefoot{
\tablefoottext{a}{Observed flux}
}
\label{XRT_1510}
\end{table*}


\section{Ultraviolet, optical, near infrared observations, data analysis, and results}

\object{PKS~1510$-$089} is included in many ongoing optical blazar monitoring programmes which provide good coverage from ultraviolet (UV) to infrared (IR) bands (Fig.~\ref{LC_uv_opt_ir}.). 
Polarimetric observations of the source were also performed. The participating observatories are described in Section 5.1-5.6 and the results of the optical observations are discussed in Section 5.7.   

\subsection{Ultraviolet and optical photometry from UVOT}
The UVOT 
covers
the 180--600\,nm wavelength range using filters: $UVW2$, $UVM2$, $UVW1$, $U$, $B$ and $V$ \citep{poole}. 
We reduced the {\it Swift}/UVOT data with the \texttt{Heasoft} 
package version 6.12 and the 20111031 release of the {\it Swift}/UVOTA CALDB. Multiple exposures in the same filter at the same epoch were summed with {\tt uvotimsum}, and aperture photometry was then performed with the task {\tt uvotsource}. Source counts were extracted from a circular region with a 5 arcsec radius centred on the source. Background counts were estimated in a surrounding annulus with inner and outer radii of 15 and 25 arcsec, respectively. The background region was selected such that it does not contain any contaminating sources.

We also compiled SEDs for all 19 epochs 
for which observations in all the six UVOT filters were available.  
The $\lambda_{\rm eff}$ and count-rate-to-flux conversion factors were derived by convolving the source spectrum with the effective areas of the UV filters.
In the same way we calculated the Galactic extinction in the various bands, using the 
\cite{cardelli} law and setting $R_V=3.1$ and $A_B=0.416$ after \cite{schlegel}.
The results were used to obtain de-reddened flux densities. Four out of the 19 SEDs (for the sake of clarity) were combined with the optical and IR data and are shown in Fig.\ref{SED_uv_opt_ir}. These epochs correspond to pre-outburst (2012 February 7, MJD~55964), two local maxima (2012 February 24, MJD~55981 and 2012 March 1, MJD~55987) and post-outburst (2012 March 26, MJD~56012) phases of the light curves.

\subsection{Optical $R$-band photometry from KVA}
\object{PKS~1510$-$089} was observed as a part of the Tuorla blazar monitoring programme{\footnote{\texttt{http://users.utu.fi/kani/1m}}}, which provides optical support observations for the MAGIC telescopes and participates in the GASP-WEBT collaboration, with the KVA 35\,cm telescope at Observatorio del Roque de los Muchachos, La Palma. The observations started on 2012 January 14 (MJD 55940) and after 2012 February 2 (MJD 55959), the source was observed every night, weather and moon conditions allowing. The data were reduced using the standard data analysis pipeline (Nilsson et al. in preparation) and the fluxes were measured with differential photometry, using the comparison stars from \citet{villata97}. 

\subsection{Optical photometry and polarisation from Steward and Perkins Observatories}
Optical (4000-7550\,\AA\,) spectropolarimetry and differential
spectrophotometry were performed at the Steward Observatory 2.3\,m Bok
Telescope using the SPOL CCD Imaging/Spectropolarimeter. These
observations were obtained as part of an ongoing monitoring programme of
$\gamma$-ray bright blazars in support of the {\it Fermi}\footnote{\texttt{http://james.as.arizona.edu/$\sim$psmith/Fermi}}.

The observations took place on 2012 January 22-29, 2012 February 13-21 and 2012 March 21-28 (MJD~55948-55955, 55970-55978, 56007-56014). The data analysis pipeline is described in \citet{smith}.   

Polarimetric and photometric $R$-band observations were
also provided by the 1.8\,m Perkins telescope of Lowell Observatory
equipped with PRISM (Perkins Reimaging System) in 2012 March. The data
analysis was done following the standard procedures as in \citet[][]{chatterjee08}.

\begin{figure*}
\includegraphics[width=0.85\textwidth, angle=270]{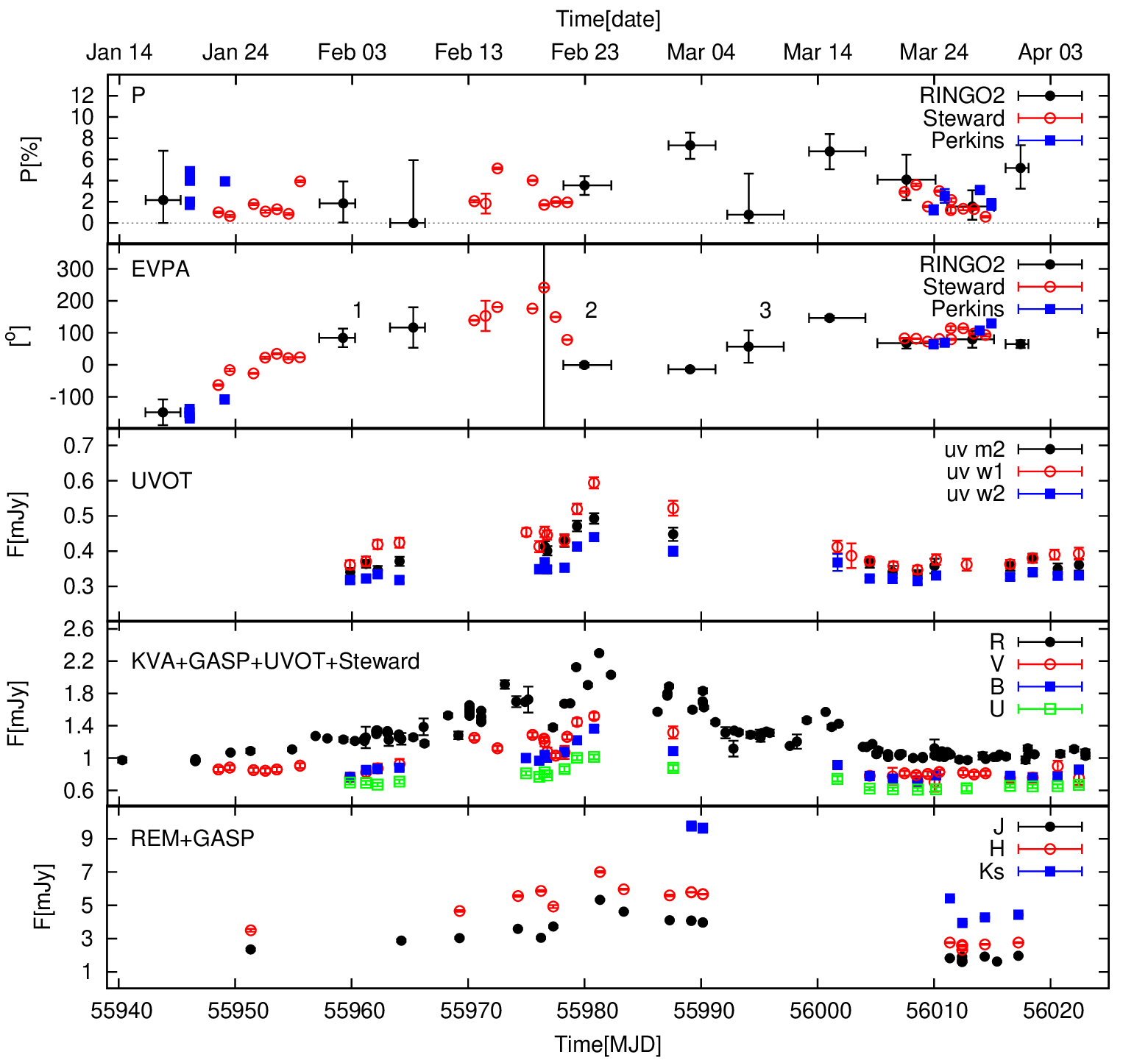}
\caption{Light curves of \object{PKS~1510$-$089} in the UV, optical and near IR bands. The optical polarisation degree and angle are also shown in the two top panels. The next panels show UV ({\it Swift}/UVOT, middle), optical (KVA/GASP/UVOT, second from bottom) and near IR (REM and GASP, bottom) light curves of the source. The numbers in the second from the top panel refers to the rotations of the EVPA discussed in the text. Vertical line indicates the time when the PA changes by $\sim90^{\circ}$ between the highlighted point and the previous point (see text). 
The fluxes are given in mJy and are not corrected for Galactic absorption.}
\label{LC_uv_opt_ir}
\end{figure*}

Because the EVPA has a $\pm180^\circ\times n$ (where $n$ = 1, 2, …)
  ambiguity, we selected the values such that the differences between
  any two points were minimised. There was one data point (Fig.~\ref{LC_uv_opt_ir}) which differed by $\sim90^\circ$ from the previous observation; we thus selected
  the EVPA for this point which does not cause a change in the direction of rotation between the two points.

\subsection{Optical and near infrared observations from GASP-WEBT}

Additional $R$-band monitoring data were collected by the GLAST-AGILE support programme (GASP) of the Whole Earth Blazar Telescope{\footnote{\texttt{http://www.oato.inaf.it/blazars/webt}}} (WEBT). 
These GASP observations of \object{PKS~1510$-$089} were performed by the following observatories: Abastumani, Calar Alto, Crimean, Lulin, Rozhen, St. Petersburg and Teide.
The source magnitude is calculated with respect to the reference stars two to six calibrated by \cite{raiteri}.
The GASP near IR data were acquired in the $J$, $H$, and $K_s$ bands with the IAC80 and Carlos Sanchez telescopes at Teide Observatory. Their calibration was performed 
using field stars with the most reliable photometry (signal to noise ratio, $\rm S/N \ga 10$ and uncertainty $\sigma < 0.11$) in the Two Micron All Sky Survey\footnote{\texttt{http://www.ipac.caltech.edu/2mass/}} (2MASS) catalogue.

\subsection{Near infrared observations from REM}

REM (Rapid Eye Mount) is a 60\,cm diameter fast reacting telescope located at La Silla, Chile. The telescope has two instruments: REMIR, an infrared imaging camera, and ROSS, a visible imager and slitless spectrograph \citep{Zer01,Chi03,Cov04a,Cov04b}. \object{PKS~1510$-$089} was observed by REM starting on 2012 January 25 (MJD~55951) during 28 nights. Typical exposure
durations were of 30\,s in the $J$, $H$, and $Ks$ filters. The data were analysed in a standard way using tools provided by the
ESO-Eclipse package \citep{Dev97}.  
Standard aperture photometry was
derived and results calibrated by a suitable number of well-exposed
2MASS objects in the field$^9$.

\subsection{Optical polarimetry observations from Liverpool Telescope}

RINGO-2 is a fast readout imaging polarimeter mounted in the fully
robotic 2-m Liverpool Telescope at Observatorio del Roque de los
Muchachos, La Palma. RINGO2 uses a hybrid $V+R$ filter, consisting of a
3mm Schott GG475 filter cemented to a 2mm KG3 filter. \object{PKS~1510$-$089} was
observed as part of a monitoring programme and started on 2012
January 19 (MJD~55945) with rather sparse sampling, but after 2012
February 21 (MJD~55978) the source was observed every night, weather
and moon conditions allowing. The data were reduced as described in
\citet{aleksic13} using a data reduction pipeline written for the
monitoring programme. Inspection of the data revealed that due
to the combination of bright moon, partial cloud coverage and low
average polarization of \object{PKS~1510$-$089}, the $S/N$ was
very low during many nights and no significant polarization was
detected. In order to improve the $S/N$ we averaged observations over
five day bins by first averaging Q/I and U/I and then computing the
unbiased degree of polarization $p_0$ and its error as in
\citet{aleksic13} with the difference that the error of EVPA was
computed using the confidence intervals in \citet{Naghizadeh-Khouei}, which are better suited for low S/N data than the $\sigma(\rm
EVPA) = 28.65*\sigma_p/p$ formula used in \citet{aleksic13}.


\subsection{Results}

The optical-UV and polarisation light curves from 2012 January to April (MJD~55952-56025) are
shown in Fig.~\ref{LC_uv_opt_ir}. The light curves
show an increasing flux peaking at near IR to UV wavelengths on 2012 February 25 (MJD~55982), the optical flux more than doubles and reaches a maximum flux of 2.23$\pm$0.39 mJy
in the R-band. After that the general trend of the light curves is decreasing. On the top of this flare, the R-band light curve which is the best sampled light curve, shows several smaller amplitude ($<$0.5\,mJy) local minima and maxima. In particular there is a dip in the light curve on 2012 February 19 (MJD~55976.5) and three local maxima after the major peak (2012 March 1, March 5 and March 13; MJD~55987, 55990 and 55999).
The flux densities varied by
5\,mJy(Ks), 1.5\,mJy (R) and 0.2\,mJy (UVW1). 
Hence, the source variability amplitude decreases as the frequency increases, as is usually found in FSRQs. This can be explained by the accretion disc emission diluting the UV emission from the jet \citep[e.g.][]{raiteri08, raiteri12} and the emission originating from the disc needs to be taken into account in the SED modelling (see Section 8). 

The optical polarisation degree was generally low ($<10^{\circ}$) during 2012 January-April compared to previous observations \citep[e.g][]{marscher}. Therefore the error bars of the measurements are rather large.  
The EVPA showed three rotations of $>180^\circ$. The first one
started in the beginning of the campaign
and ended around 2012 February 20 (MJD~55977, Fig.~\ref{LC_uv_opt_ir}). 
The rotation was $\sim$380$^\circ$, with a rotation rate of $\sim$10$^\circ$$/$day in counter-clockwise direction. The visual appearance of the
rotation curve is rather smooth, but is rather poorly sampled between January 29 and February 13 (MJD~55955 and 55970). 
The second rotation started on February 20 (MJD~55977) and ended on February 25 (MJD~55982), lasting only five days. The rotation
is $\sim250^\circ$ and the direction is opposite to the first
rotation (i.e. clockwise). After these two rotations the EVPA was
stable at $\sim$0$^\circ$ until March 7 (MJD~55993) when the third rotation started in a counter-clockwise direction and ended around March 14 (MJD~56000) at $\sim$ 150$^\circ$. On March 22 (MJD~56008) it dropped to
$\sim$ 80$^\circ$ and remained stable until the end of the
campaign.

The comparison of these rotations with the photometric
light curve and polarisation degree behaviour shows that the first rotation
takes place during an increase in the optical flux. The second rotation
starts when there is a small dip in the optical R-band light
curve and a local minimum in the polarisation degree. The rotation stops when
the optical flare peaks. The third rotation starts with a small optical
outburst and stops when the decay phase of the optical flare has reached
a plateau.
   
\begin{figure}
\includegraphics[width=0.35\textwidth, angle=270]{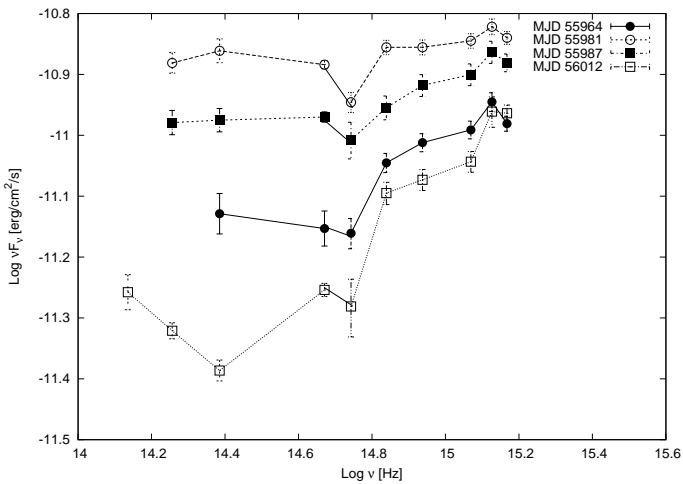}
\caption{Evolution of the infrared to ultraviolet SED from pre-outburst (MJD~55964) to two local maxima (MJD~55981 and 55987) and to post-outburst (MJD~56012) phase of the light curves. The data are corrected for Galactic absorption using \cite{schlegel}.}
\label{SED_uv_opt_ir}
\end{figure}

We constructed SEDs from IR to UV for four distinct epochs:
2012 February 7 (MJD~55964, before the outburst), February 24 (MJD~55981, peak of the outburst), March 1 (MJD~55987, second local maxima
in the R-band light curve) and March 26 (MJD~56012, quiescent state
after the outbursts), shown in Fig.~\ref{SED_uv_opt_ir}. A softening
of the SED from the pre-burst epoch to the epoch of outburst maxima is
clearly visible. In the first and last SEDs, taken before and after
the outburst, the thermal contributions from the accretion disc are
again clearly visible as a strong increasing trend in the optical and
UV bands. This behaviour was also seen for the 2009 outburst reported
in \citet{dammando11}.

\section{Radio observations, data analysis, and results}

\object{PKS~1510$-$089} is part of the numerous blazar radio monitoring programmes extending from 2.6\,GHz to 230\, GHz by F-GAMMA, Medicina, UMRAO, OVRO, Mets\"ahovi, VLBA and the Submillimeter Array. The observations collected for this paper are presented in Sections 6.1-6.7 and the results discussed in 6.8. 

\begin{figure*}
\includegraphics[width=0.41\textwidth, angle=270]{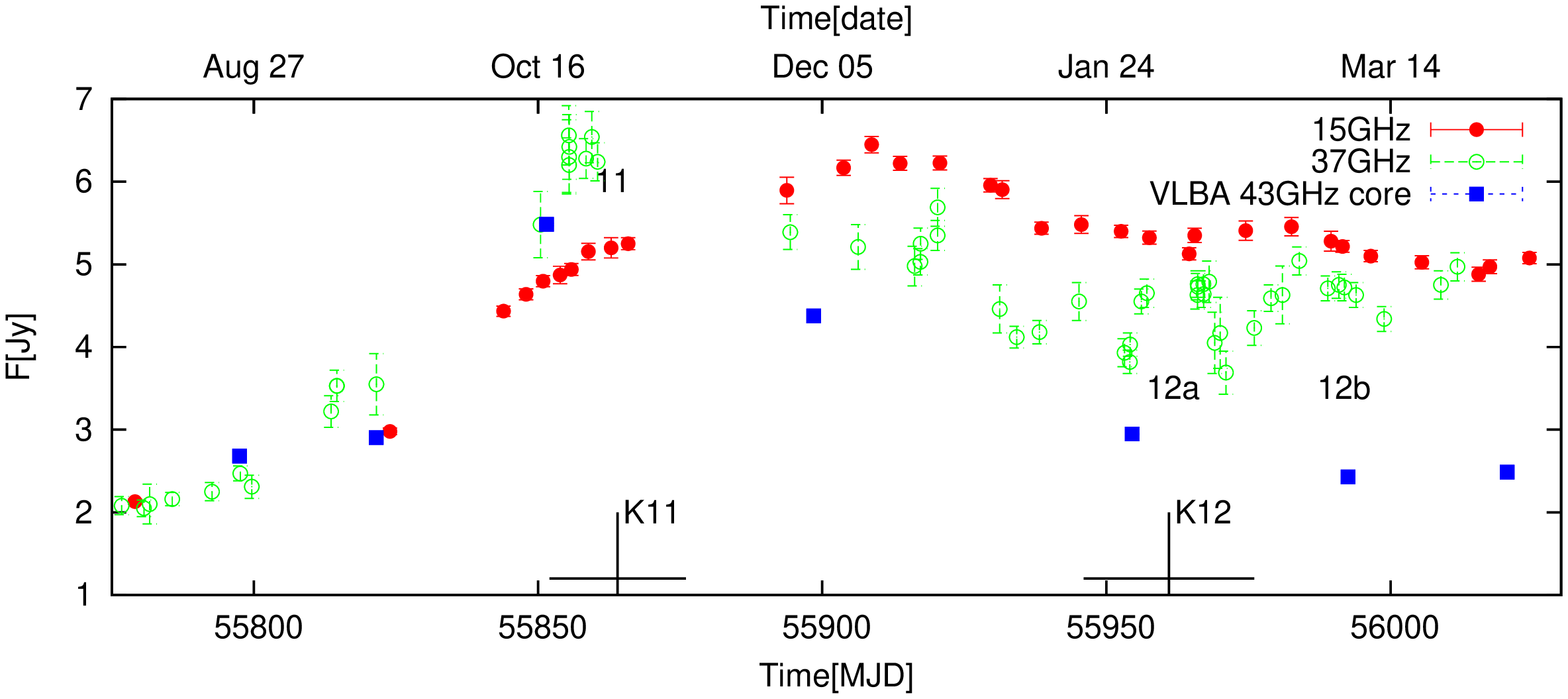}
\caption{15\,GHz, 37\,GHz and 43\,GHz VLBA core long-term light curves from MJD~55750 (2011 July 8) to MJD~56030 (2012 April 13). The flux of the VLBA core at 43\,GHz traces the shape of the 37\,GHz light curve, 
indicating that the major part of the total flux originates in there. Moreover, the new components found at 43\,GHz are coincident with flux increase in the 37 GHz band. The symbols at the bottom of the plot show the zero separation epochs with the error bars of the components K11 and K12 from the 43\,GHz VLBA core (see text).}
\label{radio_long}
\end{figure*}

\subsection{Submillimeter Array}
The 230\,GHz (1.3\,mm) light curve was obtained at the Submillimeter Array (SMA) on Mauna Kea (Hawaii). The SMA is an 8-element interferometer, consisting of 6\,m dishes that may be arranged into configurations with baselines as long as 509\,m, producing a synthesised beam of sub-arcsecond width. \object{PKS~1510$-$089} is included in an ongoing monitoring programme at the SMA to determine the fluxes of compact extragalactic radio sources that can be used as calibrators at mm wavelengths \citep{gurwell07}. Observations of available potential calibrators are usually observed for three to five minutes, and the measured source signal strength calibrated against known standards, typically solar system objects (Titan, Uranus, Neptune, or Callisto). Data from this programme are updated regularly and are available at the SMA website\footnote{\texttt{http://sma1.sma.hawaii.edu/callist/callist.html}}.

\subsection{Mets\"ahovi Radio Telescope}
The 37 GHz observations were made with the 13.7 m diameter Mets\"ahovi radio telescope, which is a radome enclosed paraboloid antenna situated in Finland. The measurements were made with a 1 GHz-band dual beam receiver centred at 36.8\,GHz. The beamwidth is 2.4 arcmin. The high electron mobility pseudomorphic transistor front end operates at room temperature. The observations were performed in an ON--ON configuration alternating the source in each feed horn, with the second horn observing the sky. The flux density scale was set by observations of calibrator DR~21. The sources NGC 7027, 3C 274 and 3C 84 were used as secondary calibrators. A detailed description of the data reduction and analysis is given in \citet{Metsaehovi}. The error estimate in the flux density includes contributions from the measurement rms and the uncertainty of the absolute calibration. 
The \object{PKS~1510$-$089} observations were done as part of the regular monitoring programme and the GASP-WEBT campaign.

\subsection{Owens Valley Radio Observatory}
Regular 15 GHz observations of \object{PKS~1510$-$089} were carried out as part
of a high-cadence $\gamma$-ray blazar monitoring programme using the
Owens Valley Radio Observatory (OVRO) 40 m telescope in Owens Valley,
California \citep{Richards11}. This programme, which commenced in late
2007, now includes about 1800 sources, each observed with a nominal
twice per week cadence.

The OVRO 40 m uses off-axis dual-beam optics and a cryogenic high
electron mobility transistor low-noise amplifier with a 15.0\,GHz
centre frequency and 3\,GHz bandwidth. The telescope and receiver
combination produces a pair of approximately Gaussian beams (157
arcsec full width half maximum; FWHM), separated in azimuth by 12.95 arcmins.  The total system
noise temperature is about 52 K, including receiver, atmosphere,
ground, and CMB contributions. The two sky beams were Dicke switched
using the OFF-source beam as a reference, and the source is alternated
between the two beams in an ON-ON fashion to remove atmospheric and
ground contamination. A noise level of approximately 3--4\,mJy in
quadrature with about 2\% additional uncertainty, mostly due to
pointing errors, is achieved in a 70\,s integration period.
Calibrations were performed using a temperature-stable diode noise
source to remove receiver gain drifts and the flux density scale was
derived from observations of 3C~286 assuming the \citet{Baars77} value
of 3.44\,Jy at 15.0\,GHz. The systematic uncertainty of about 5\% in
the flux density scale is not included in the error bars. Complete
details of the reduction and calibration procedure are found in
\citet{Richards11}.

\begin{figure*}
\includegraphics[width=0.75\textwidth, angle=270]{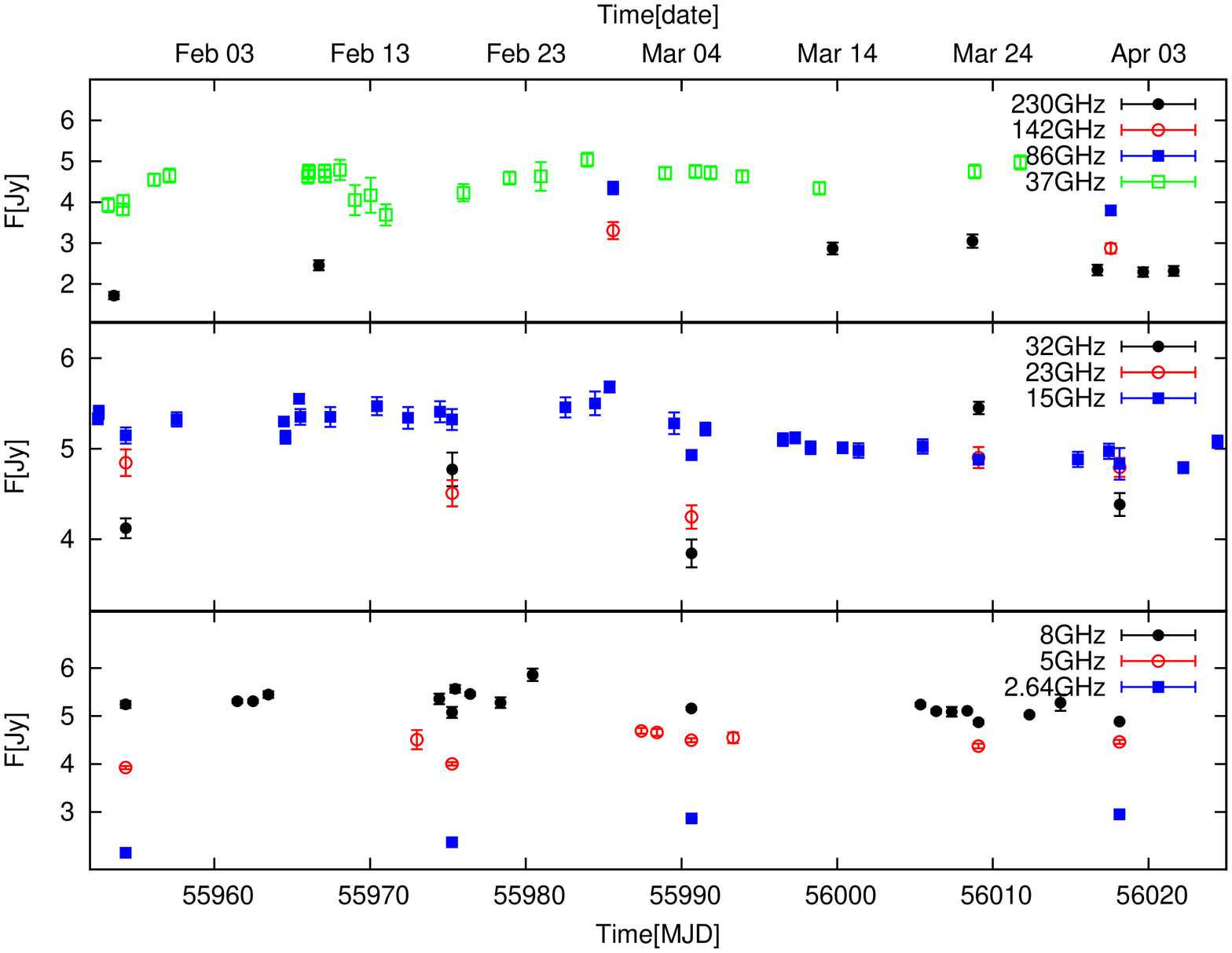}
\caption{High frequency (top), medium frequency (middle) and low frequency (bottom) light curves from SMA, Mets\"ahovi, OVRO, UMRAO, Medicina and F-GAMMA programme for the campaign period.}
\label{radio_short}
\end{figure*}

\subsection{F-GAMMA programme}
The cm/mm radio light curves of PKS\,1510--089 have been obtained within the framework of a
{\sl Fermi}-GST related monitoring programme of $\gamma$-ray blazars 
\citep[F-GAMMA programme,][]{2007AIPC..921..249F,2008arXiv0809.3912A}. 
 The overall frequency range spans from 2.64\,GHz to 142\,GHz using the 100\,m radio telescope located in Effelsberg, Germany and IRAM 30\,m located on Pico Veleta in the Spanish Sierra Nevada.

The Effelsberg measurements were conducted with the secondary focus heterodyne receivers
at 2.64, 4.85, 8.35, 10.45, 14.60, 23.05, 32.00 and 43.00\,GHz. The observations were performed
quasi-simultaneously with cross-scans, slewing in
azimuth and elevation across the source position with an adaptive number of sub-scans until the desired
sensitivity is reached \citep[for details, see][]{2008A&A...490.1019F, 2008arXiv0809.3912A}. Consequently,
pointing off-set correction, gain correction, atmospheric opacity correction and
sensitivity correction have been applied to the data.

The IRAM 30\,m observations were carried out with calibrated
cross-scans using the new EMIR horizontal and vertical polarisation
receivers operating at 86.2 and 142.3\,GHz. The opacity corrected
intensities were converted into the standard temperature scale and
finally corrected for small remaining pointing offsets and systematic
gain-elevation effects. The conversion to the standard flux density
scale was done using the instantaneous conversion factors derived from
frequently observed primary (Mars, Uranus) and secondary (W3(OH),
K3-50A, NGC\,7027) calibrators.

\subsection{UMRAO}
Centimetre band total flux density observations were obtained with the
University of Michigan Radio Observatory (UMRAO) 26\,m paraboloid
located in Dexter, Michigan, USA. The instrument is equipped with
transistor-based radiometres operating at frequencies centred at 4.8,
8.0, and 14.5\,GHz with bandwidths of 0.68, 0.79, and 1.68\,GHz,
respectively. Dual horn feed systems are used at 8 and 14.5 GHz, while
at 4.8 GHz a single-horn, mode-switching receiver is employed. Each
observation consisted of a series of 8 to 16 individual measurements
over approximately a 25 to 45 minutes time period, utilizing an ON-OFF
observing technique at 4.8 GHz, and an ON-ON technique (switching the
target source between the two feed horns which are closely spaced on
the sky) at 8.0 and 14.5\,GHz. As part of the observing procedure,
drift scans were made across strong sources to verify the telescope
pointing correction curves, and observations of nearby calibrators
were obtained every one to two hours to correct for temporal changes in
the antenna aperture efficiency. The \object{PKS~1510$-$089} observations were done as part of the regular monitoring programme and the GASP-WEBT campaign.

\subsection{Medicina}
The Medicina telescope is a 32\,m parabolic antenna located 30\,km from Bologna, Italy, performing observations at both 5 and 8.4 GHz\footnote{\texttt{http://www.med.ira.inaf.it/parabola\_page\_EN.htm}}. FWHM beamwidth is 38.7 arcmin/frequency (GHz). 
 We used the new enhanced single-dish control
system acquisition system, which provides enhanced sensitivity and
supports observations with the cross scan technique. All observations were
performed at both 5 and 8.4 GHz; the typical on-source time is 1.5 minutes
and the flux density was calibrated with respect to 3C 286. PKS~1510$-$089 was observed
during 2012 January-April as part of the regular monitoring programme and the GASP-WEBT campaign.

\begin{figure}
\includegraphics[width=0.35\textwidth, angle=270]{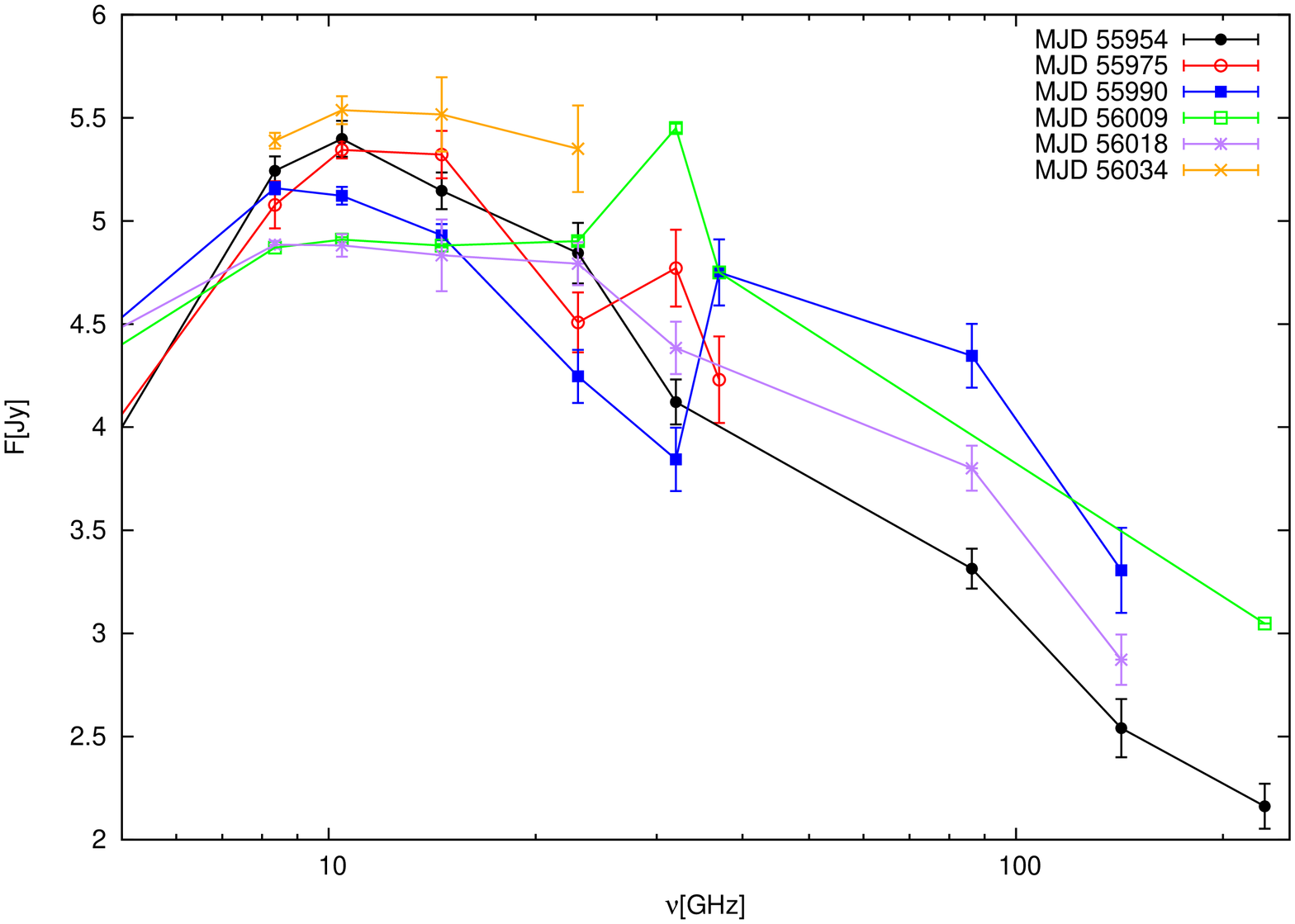}
\caption{Evolution of the radio spectra over the campaign period from 2012 January 28 to April 17 (MJD~55954, 55975, 55990, 56009, 56018 and 56034). The first radio outburst 11 in Fig.~\ref{radio_long} dominates the spectra in the first epoch, while in the second epoch new outburst 12a in Fig.~\ref{radio_long} is apparent in the high frequencies. In the third epoch 12b becomes visible in the highest frequencies with the peak moving to lower energies in the fourth and fifth epochs.}
\label{radio_sed}
\end{figure}

\begin{figure}
\includegraphics[width=0.49\textwidth]{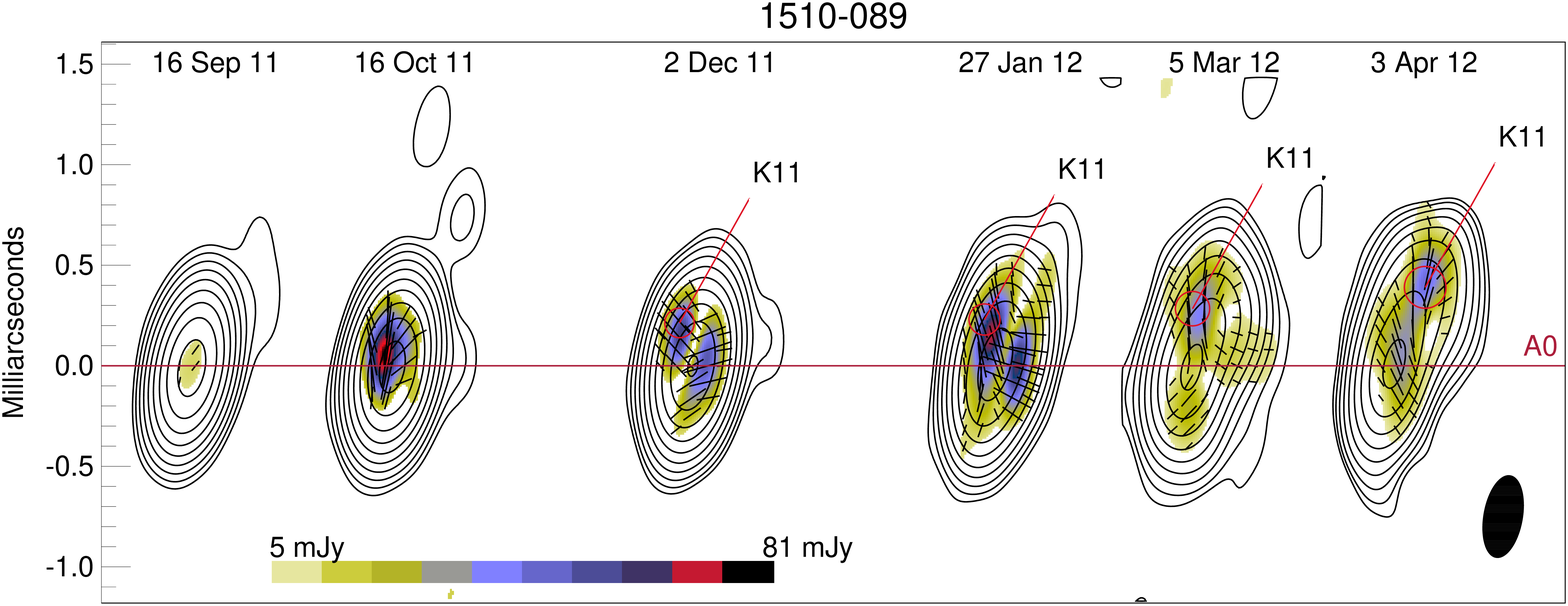}
\hspace{10 mm}
\includegraphics[width=0.49\textwidth]{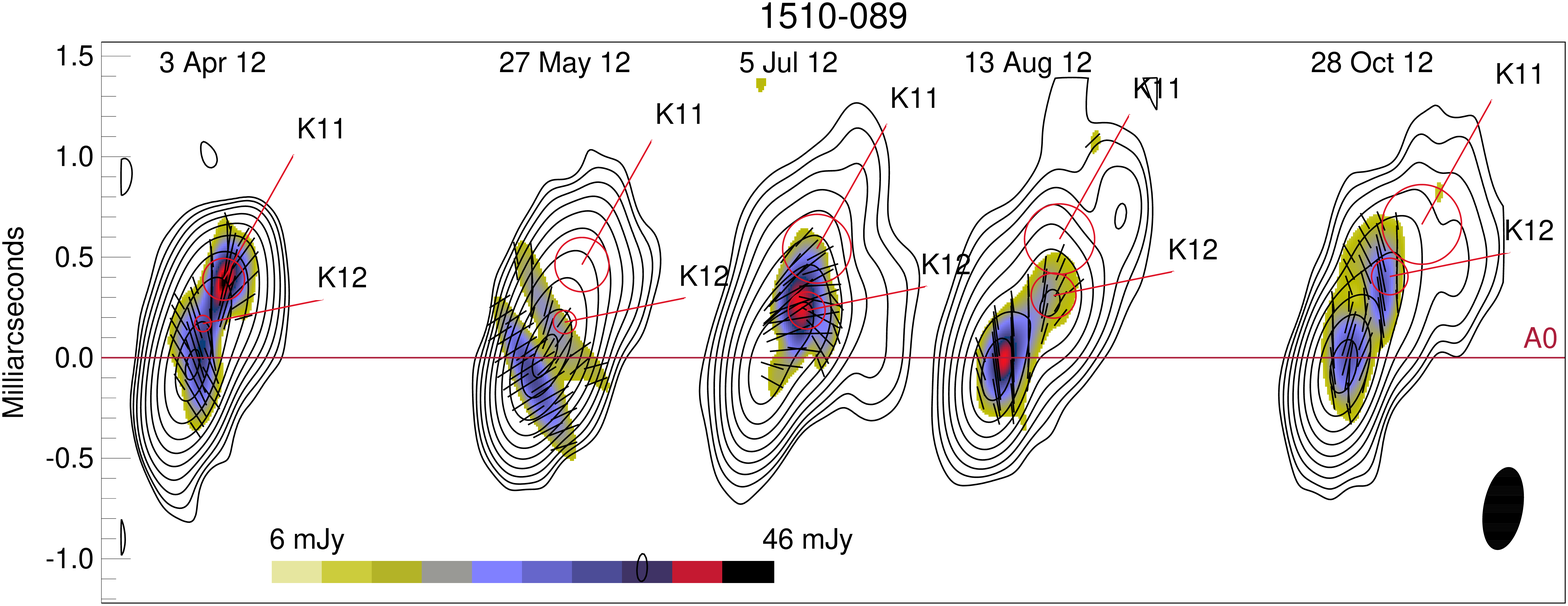}
\caption{43\,GHz total (contours) and polarised (colour scale) intensity images of PKS~1510$-$089
from  2011 September to 2012 April (top) and 2012 April to October (bottom)
with $S_{\rm peak}$=2.58~Jy/beam,  $S_{\rm peak}^{\rm pol}$=46~mJy/beam, and a Gaussian restoring beam=0.14$\times$0.39~mas$^2$ at $PA$=-10$^\circ$
(in the right bottom corner); contours represent 0.25, 0.5,...,64, 90\% of the peak intensity; line segments within the colour scale show direction of linear polarisation; solid lines indicate position of components across the epoch, the core A0, knot K11, and knot K12.} 
\label{k12}
\end{figure}

\subsection{Very Long Baseline Array}
VLBA is a system of ten radio-telescope antennas, each with a dish 25\,m in diameter located from Mauna Kea in Hawaii to St. Croix in the U.S. Virgin Islands.  
VLBA observations were performed as a part of the
Boston University $\gamma$-ray blazar
monitoring programme at 43\,GHz. The observations were carried out with the VLBA recording system using eight 8\,MHz wide channels, each in right
and left circular polarization, with 15$-$20 scans of three to five minutes
duration. All ten antennas were used except at epochs affected by weather or receiver failure.
The observations are performed about once per month. The data were reduced and modelled in the same manner as described in \citet{jorstad05,jorstad07}. In short: the initial correlation was carried out at the National Radio Astronomy
Observatory (NRAO) Array Operations Center in Socorro, New
Mexico and subsequent calibration was performed with the astronomical image processing system (AIPS) software supplied
by NRAO, while images were made with the Caltech software
DIFMAP. These calibrations included application
of the nominal antenna-based gain curves and system temperatures, as well as correction for sky opacity, followed by iterative
imaging plus phase and amplitude self-calibration. The flux-density correction factors from \citet{jorstad05} were used for the final adjustment of the flux-density scale in the images. 
In addition to the kinematics of the jet, the total polarisation data and the polarisation of the VLBA core were also analysed. Also these analysis followed the methods in \citet{jorstad05}.

\subsection{Results}

In second half of 2011 \object{PKS~1510$-$089} showed extremely high cm- and mm-band radio
flux \citep{fgamma-atel, medicina-atel,brazil-atel}. During the outburst the flux increased from 2\,Jy to 7\,Jy. The outburst 
peaked first at higher frequencies, the peak at 37\,GHz was reached
around 2011 October 21 (MJD~55855) and at 15\,GHz on $\sim$ 2011
December 15 (MJD~55910, see Fig.~\ref{radio_long}, outburst ``11''). After the maximum
was reached the two radio light curves showed decreasing
flux. However, there are several smaller amplitude outbursts (amplitudes 1-2\,Jy) visible in the
both light curves peaking at 2012 January 20 and February 25 (MJD~55946 and 55982) at 15\,GHz. The last outburst at 15\,GHz appears to
be a sum of two outbursts seen at 37\,GHz peaking at 2012 February 8 and February 25 (MJD~55965 and 55982, outburst ``12a'' and ``12b'' in Fig.~\ref{radio_long}).

Figure~\ref{radio_short} shows radio light curves from all frequencies
from the observing campaign period. In the lowest frequencies
(2-8\,GHz) there is very little variability
while at higher frequencies variability is clearly present at all
frequencies, but the rather sparse sampling does not allow us to identify
outbursts from other than 15\,GHz and 37\,GHz light curves.

The radio spectral evolution from 2012 January 28 to April 17 (MJD~55954 to 56034) is shown in Fig.~\ref{radio_sed}. In the four first spectra at low
frequencies the dominating component is the decaying major outburst. At higher frequencies the new outburst 12a is first visible on 2012 February 18 (MJD~55975). The outburst 12b is first visible on 2012 March 4 (MJD 55990) and the peak then moves to lower frequencies. In the
last two spectra this outburst is visible as a flattening of the spectra above 15\,GHz and increased flux. Both outbursts follow the typical spectral evolution of radio outbursts. In the initial (growth) stage, the synchrotron self-absorption turnover frequency decreases and the turnover flux density increases. In the second (plateau) stage, the turnover frequency decreases while the turnover flux density remains roughly constant. During the third (decay) stage both turnover frequency and flux density decrease.
The behaviour is in agreement with the three stage evolution of the
shock-in-jet model of \citet{MG85}; in the first stage the inverse Compton losses dominate, in the second the synchrotron losses and in the third the adiabatic losses.

\begin{figure}
\includegraphics[width=0.24\textwidth,angle=270]{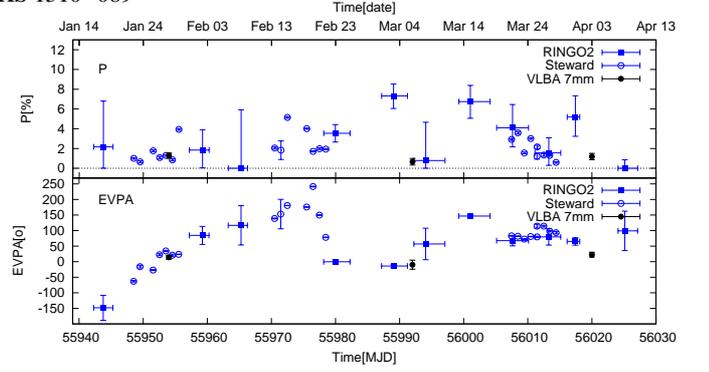}
\caption{Radio and optical polarisation behaviour of PKS~1510$-$089 in 2012 February-April.} 
\label{pol}
\end{figure}

The VLBA 43\,GHz images reveal a new component (named K11)
corresponding to the major radio outburst of the second half of 2011
appearing in 2011 December as already reported in \citet{orienti}
using the MOJAVE 15\,GHz data (see Fig.~\ref{k12}). The apparent speed
of the component, $(19.34\pm1.85)$c, and the zero separation epoch
2011 October 29 (MJD~55864$\pm$12) 
ones derived by \citet{orienti}. In 2012 April there was a second new
component appearing in the images (named K12). It had an apparent
speed of $(16.26\pm2.43)$c and a time of ejection of 2012 February 3
(MJD~55961$\pm$15) 
55961$\pm$15days).  The zero separation epochs of these components
agree very well with the local maxima in the 37\,GHz light curve
according to the general trend found in \citet{savolainen}. The VLBA
polarisation data showed in general a rather low polarisation of the
core (1-3\%) compared to the historical values from \citet{jorstad07}
(0.9-9.7\%). The observed EVPA of the core between 2012 January and
April was between -10$^o$ and 25$^o$. The sparse sampling does not
allow us to trace possible rotations of the EVPA, but as shown in
Fig.~\ref{pol}, the EVPA values of the VLBA core seem to trace 
close those of the optical emission.

\section{Multi-Frequency light curves}

\begin{figure*}
\includegraphics[width=1.05\textwidth, angle=270]{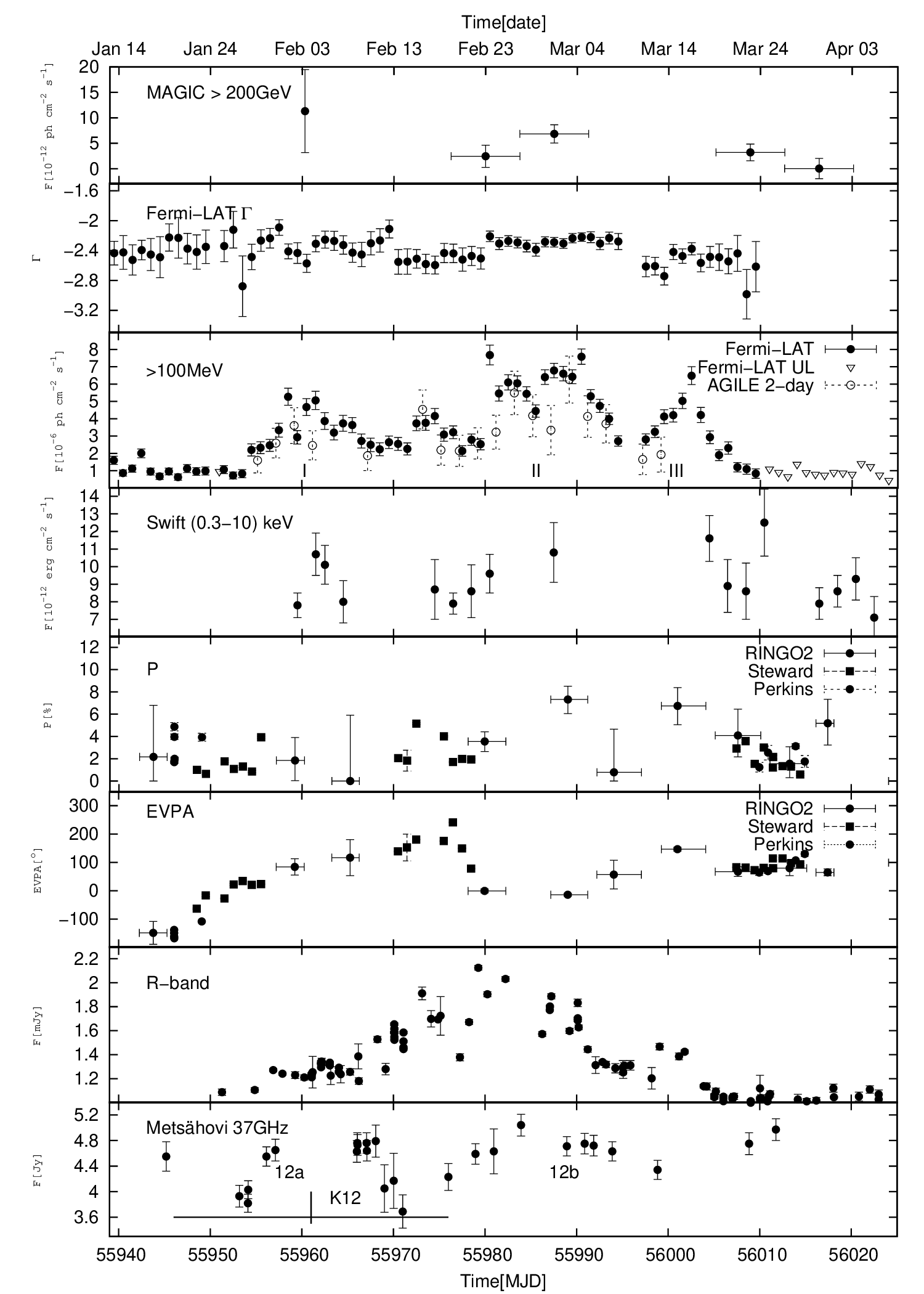}
\caption{Multi-frequency light curve of PKS~1510$-$089 from VHE $\gamma$ rays to radio in February-April 2012. The symbol marked K12 in the bottom panel shows the zero separation epoch of the VLBA component K12 (see Section 6.8) from the 43\,GHz VLBA core. The numbering of the HE $\gamma$-ray and 37\,GHz radio flares is described in the text.}
\label{MWL_lc}
\end{figure*}

Figure~\ref{MWL_lc} shows the MAGIC,
{\it Fermi}-LAT, AGILE-GRID, {\it Swift}, optical polarisation, $R$-band
photometry and 37\,GHz light curves of \object{PKS~1510$-$089} in 2012 February-April. The {\it Fermi}-LAT light curve showed three distinct flares with flux increase more than factor five compared to quiescent state flux: flare I (2012 January 29 to February 13, MJD$\sim$55955-55970), flare II (2012 February 23 to March 9 , MJD$\sim$55980-55995) and flare III (2012 March 14 to March 19, MJD$\sim$56000-56005). Additionally there was a smaller amplitude ($\sim$ factor four) flare between flare I and II. The first two flares also triggered the AGILE alert system,
and are evident in the two day AGILE-GRID light curve,
while during flare III the source gradually exited the AGILE
field of view. As discussed in Section 3 AGILE and {\it Fermi}-LAT data hint for a marginal harder when brighter trend.
During these flares the VHE $\gamma$-ray flux remained rather unchanged. 
The flares were all characterised by different multi-frequency behaviour at lower energies. The first $\gamma$-ray outburst coincided with an X-ray peak. The first and second $\gamma$-ray flares were accompanied by quasi simultaneous flares in 37\,GHz radio (flare I in $\gamma$ rays is simultaneous to flare 12a in radio and flare II in $\gamma$ ray is simultaneous to flare 12b in radio, see Fig.~\ref{MWL_lc}). During the first outburst there was also a rotation of the EVPA
of $>180^\circ$. This outburst also coincided with the zero
separation epoch of new knot from the 43\,GHz VLBA core (see Section 6.8).

During the second $\gamma$-ray flare there was an optical outburst
and in the very beginning a second rotation of the EVPA with $>180^\circ$, but this rotation had a very short duration and it was in the opposite
 direction from the first one. During this rotation there was
also a local minimum of the polarisation degree, and this rotation looks
very similar to the one observed in 3C~279 during the $\gamma$-ray event seen by {\it Fermi}-LAT in 2009 April \citep{nature}. However, while the optical flux started to decrease, the $\gamma$-ray flare continued and the
optical polarisation degree started to increase.

After these events the EVPA stayed constant until the third
rotation started apparently simultaneously with the third outburst in
the $\gamma$-ray light curve. During the outburst the
degree of polarisation stayed constant. There was a gap in the 37\,GHz
light curve, however the emission level was similar before and after the gap.

The overall outbursting event had several similarities to the $\gamma$-ray
flaring event in 2009 discussed in \cite{marscher,abdo10,dammando11}:
ejection of the knot from the VLBA core, accompanied activity in the
millimetre wavelengths and the rotation of the optical polarisation
angle. However, there are also some differences: there was no preceding
$\gamma$-ray flare, but the activity in radio and $\gamma$ rays started
simultaneously. Also the observed rotation of the optical
polarisation angle was shorter in duration ($\sim$ 30 days) and the
rotation was only $\sim$ 380$^\circ$ instead of $>720^\circ$ seen in
2009.

\cite{marscher} interpreted the 2009 outburst in terms of the
phenomenological model presented for BL Lacertae in
\cite{Marscher_nature}. In this model the rotation of the
polarisation angle is caused by a moving emission feature following a
spiral path as it propagates through the toroidal magnetic field of
the acceleration and collimation zones. The emission feature passes the
43\,GHz VLBA core, interpreted as a standing conical shock
\citep{Marscher_nature}, which compresses the knot. The synchrotron
flares occur when the energisation of the electrons increases suddenly
while the $\gamma$-ray flares with very weak optical counterparts are
produced by an increase of the local seed photon field in
optical and IR wavelengths. 
The same scheme can be adopted to the multi-frequency light curve discussed here:
flare I takes place as the emission feature passes the core while flare II is
caused by the sudden energisation of the electrons of the emission
feature and flare III by the sudden increase in the local seed photon
field. 

Unlike the millimetre, optical and HE $\gamma$ rays, the X-rays did not
show strong variability. The X-ray light curve showed a general shape similar to that of the HE $\gamma$-ray light curve. 
However, the sparse sampling and the small amplitude of variability in the X-ray light curve, do not allow us to draw a strong conclusion on the connection. The X-ray spectrum was hard, as in the previous observations \citep{dammando11,kataoka08}, which is a signature of a hard electron population
with slope 1.6-2.0. The observed properties are in agreement with the
conclusion of \cite{kataoka08} that the X-ray spectrum is a result of
Comptonization of IR radiation produced by hot dust located in
the surrounding molecular torus. Direct mid-IR (3.6-- 160$\mu$m) observations give an upper limit on the luminosity from thermal emission from such dust to be $2.3\cdot10^{45}$ erg s$^{-1}$ \citep{malmrose}.

As discussed in Section 5.7, the behaviour of the optical polarisation
EVPA during the observing campaign was
particularly interesting showing three distinct rotations of $>180^\circ$. In addition to \object{PKS~1510$-$089} \citep{marscher} and BL~Lac
\citep{Marscher_nature} such rotations have been reported for 3C~279
in coincidence with $\gamma$-ray flaring events
\citep{larionov,nature,3C279,aleksic13}. In these papers the rotations
have been interpreted as a signature of the geometry, in particular as caused by a
bent trajectory that the moving emission feature is following. For 3C~279 the rotations have been changing direction
between epochs, which was interpreted as a signature of an actual
bend in the jet \citep{nature,nalewajko10}. However, here the second rotation was very fast, the rotations took place very close in time and the multi-frequency data
suggests that a major part of the emission would originate in one
single emission region. Therefore, a bend does not appear a likely
explanation for the change of the direction of the rotations.

As discussed in \citet[][and references therein]{marscher}, the
rotations can also be explained by a turbulent magnetic field within
the emission region where cells with random magnetic field
orientations enter and exit the emission region causing a random walk
of the resultant polarisation vector and apparent rotation. For
rotations caused by the turbulent magnetic field both directions
should be as likely and they should occur at random
times. Additionally the appearance of the rotations caused by
turbulence is not very smooth. Turbulence as a possible cause of the
rotations is favoured by the fact that the rotations with different
directions took place very close in time. According to
\citet{marscher14} such rotations are expected when turbulent plasma
flows at relativistic speeds down a jet and crosses the standing
shock.

In summary we conclude that the multi-frequency light curves show
compelling evidence that the emission during this flaring epoch is
dominated by a moving emission feature located close to the VLBA
43\,GHz core. As described before such evidence has been found for \object{PKS~1510$-$089} as well as in several other sources \citep[e.g.][and references therein]{jorstad13}. The complicated flaring pattern, showing variable
synchrotron to Compton ratios and different time scales in diffent
wavebands, suggest that additionally both the emission region and the
underlying jet might have some substructures.

\section{Spectral energy distribution}

We construct the SED for \object{PKS~1510$-$089} combining the radio data from
F-GAMMA and Mets\"ahovi with infrared, optical and UV data
from REM, GASP-WEBT and UVOT, X-ray data from {\it Swift}-XRT and
$\gamma$-ray data from {\it Fermi}-LAT and MAGIC. The radio to X-ray
data are quasi-simultaneous taken from 2012 March 1 to March 4
(MJD~55987-55990) while the {\it Fermi}-LAT data cover the main MAGIC
observation periods (2012 February 19 to March 5 and March 15 to April
3, MJD~55976-55991 and MJD~56001-56020) and AGILE data the period from
2012 February 20 to March 5 (MJD~55977.5-55991.5).

\begin{figure}
\includegraphics[width=0.45\textwidth]{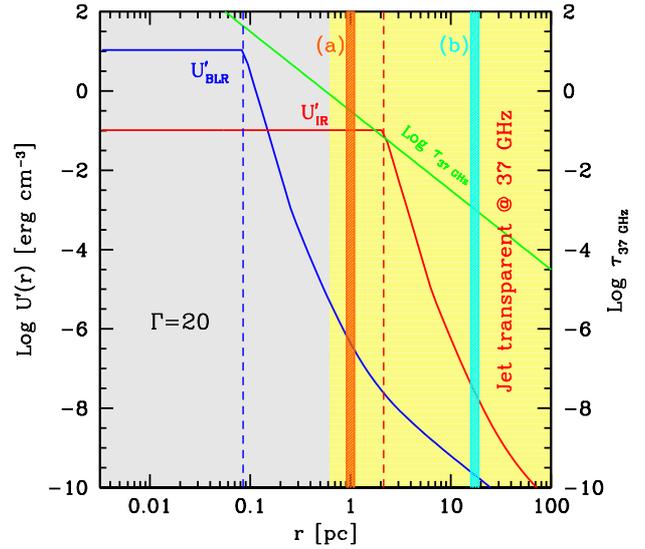}
\caption{Energy density of the photon field as function of the distance from the central engine. The blue lines refer to the BLR and the red lines to infrared torus. The green line indicates the $\tau_{37 GHz}$ (calculated using the magnetic field derived for case {\it a} model, see text) and yellow zone indicates the area at which the jet is transparent at 37\,GHz. The dashed lines indicate the assumed size of BLR (blue) and dust torus (orange). The thick red and cyan vertical lines indicate the regions which we selected for our SED modelling.}
\label{u}
\end{figure}

The SEDs of FSRQs are conventionally modelled with a small emission region (typically size of $\sim$10$^{16}$ cm) close to the central engine, in regions where the dense radiation field generated by the direct and reprocessed 
accretion disc emission is thought to provide the ideal environment for efficient IC emission \citep{dermer, sikora94}. 
There is, however, growing evidence that, at least in some objects and/or at some epochs, the emission could occur far downstream in the jet \citep{sikora08,Marscher_nature,1222}.

For \object{PKS~1510$-$089} at the epoch analysed here, the multi-frequency light curves and the ejection of a new component from the 43\,GHz VLBA core 
point to the co-spatial siting of the $\gamma$-ray and millimetre flaring activity. Since the inner regions of the jet are highly opaque to low frequency photons through synchrotron self-absorption \citep[as indeed observed for the great majority of FSRQs, e.g.][]{giommi2012}, the $\gamma$-ray and millimetre emission region has to be located farther out in the jet, at distances at which the jet is transparent at radio frequencies. Another compelling indication that the $\gamma$ rays are not produced very close to the nucleus is that, in this case, one would expect a strong depression of the emission above $\sim20$ GeV due to absorption through interactions with the UV-optical emission of the BLR clouds \citep[e.g.][]{donea, sitarek, FM, PoutanenStern}. 
Instead, the combined {\it Fermi}-LAT and MAGIC $\gamma$-ray spectrum does not show signatures of strong absorption, but a smooth log parabola shape. This is similar to what was observed for PKS~1222+216 \citep{1222} and for other few other FSRQs whose LAT spectrum extends well above 10-20\,GeV, supporting the idea of emission occurring beyond the BLR radius \citep{pacciani,tavecchio13}.

The simultaneous millimetre and $\gamma$-ray 
light curves show similar variability patterns on a weekly time scales, and are therefore consistent with a large dominating emission region, $R\sim ct_{\rm var} \delta = 2\times 10^{17} (\delta/10)$ cm, where $\delta$ is the relativistic Doppler factor. The low compactness implied by such large dimensions makes the synchrotron self-Compton process, in which the seed photons are produced in the jet via synchrotron radiation \citep[e.g.][]{maraschi92}, highly inefficient. It is thus unable to produce the observed $\gamma$-ray emission \citep[see e.g.][for the case of 3C 279]{lindfors05}, and therefore the seed photons for IC scattering must be provided by some external field.

The radiative environment for the jet in \object{PKS~1510$-$089} is schematically described in Fig.~\ref{u}, which reports the energy density of the external radiation in the jet co-moving frame as a function of the distance from the central engine. Two components are considered, namely the emission of the BLR clouds in the innermost regions (blue), and the contribution provided by the thermal emission of dust organised in the molecular torus, at larger scales (red).
The external energy density is assumed to be constant within 
the corresponding radius of the emitting structure, $r_{{\mathrm BLR}}$ and $r_{IR}$, for the BLR and the IR torus respectively, and shows a rapid decline beyond it. The detailed geometry and extension of the BLR and of the IR torus are still under debate, but values typically adopted for the extensions are of order $0.1-1$ parsec and $1-5$ parsecs, respectively. \cite{nalewajko12} estimated that for
\object{PKS~1510$-$089} these values are 
$r_{{\mathrm BLR}}=0.07$ pc and $r_{{\mathrm Torus}}=3.2$ pc. 
The curves in Fig.~\ref{u} have been calculated following \citet{ghisellini09}, who provide simple scaling laws for the dimensions of the BLR and the IR torus, depending only on the accretion disc luminosity, $L_{\rm disc}$. In the literature there are several estimates of the disc luminosity \citep{celotti,nalewajko12},
all in the range $3-7\times 10^{45}$ erg s$^{-1}$. In the following we assume $L_{\rm disc}=6.7\times 10^{45}$ erg s$^{-1}$, as inferred from the observed ``blue bump'' traced by UVOT. 
With the adopted $L_{\rm disc}$ the estimates of \citet{ghisellini09}
provide $r_{{\mathrm BLR}}=0.086$ pc and $r_{{\mathrm Torus}}=2.15$ pc.
The resulting $L_{\rm Torus}$ is compatible with the upper limits from mid-IR observations \citep{malmrose}.

We reproduced the observed SED by assuming that the emission region (blob) is filled with electrons following a smoothed broken power-law energy distribution with normalization $K$ between $\gamma _{\rm min}$ and $\gamma _{\rm max}$, with slopes $n_1$ and $n_2$ below and above the break at $\gamma _{\rm b}$ as in \citet{tavecchio98} and \citet{maraschi03}. We assume a conical geometry for the jet, characterised by a semi-aperture angle $\theta_{\rm j}=0.1^\circ$ \citep{jorstad05}.
Electrons emit through synchrotron and IC mechanisms. 
The relative luminosities of the IC components arising from the different target photon populations are proportional to the level of the corresponding energy densities (measured in the jet frame), when the scattering takes place in Thomson regime. The energy density, in turn, depends on the distance along the jet as reported in Fig.~\ref{u}. 


\begin{figure*}
\includegraphics[width=0.45\textwidth]{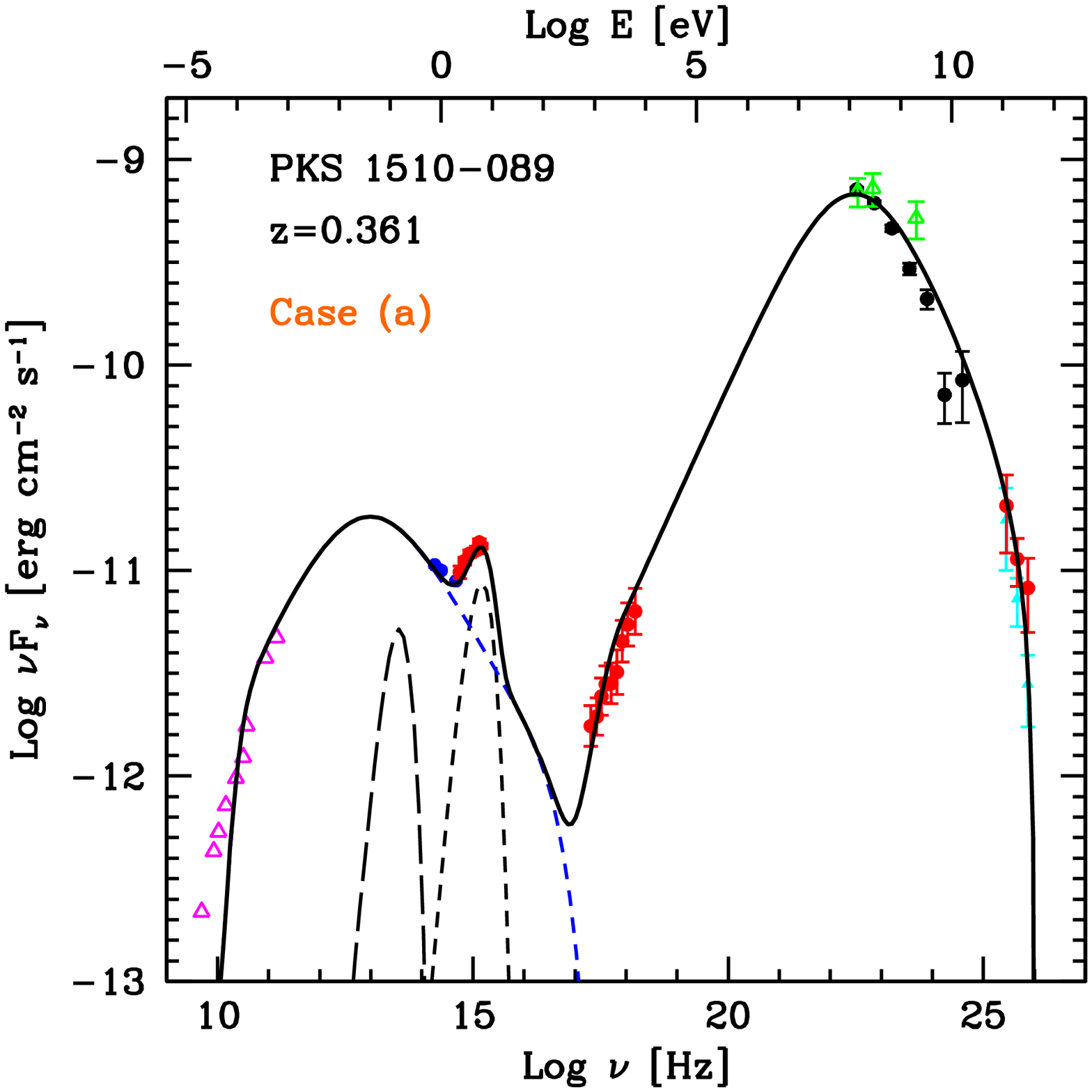}
\hspace{10 mm}
\includegraphics[width=0.45\textwidth]{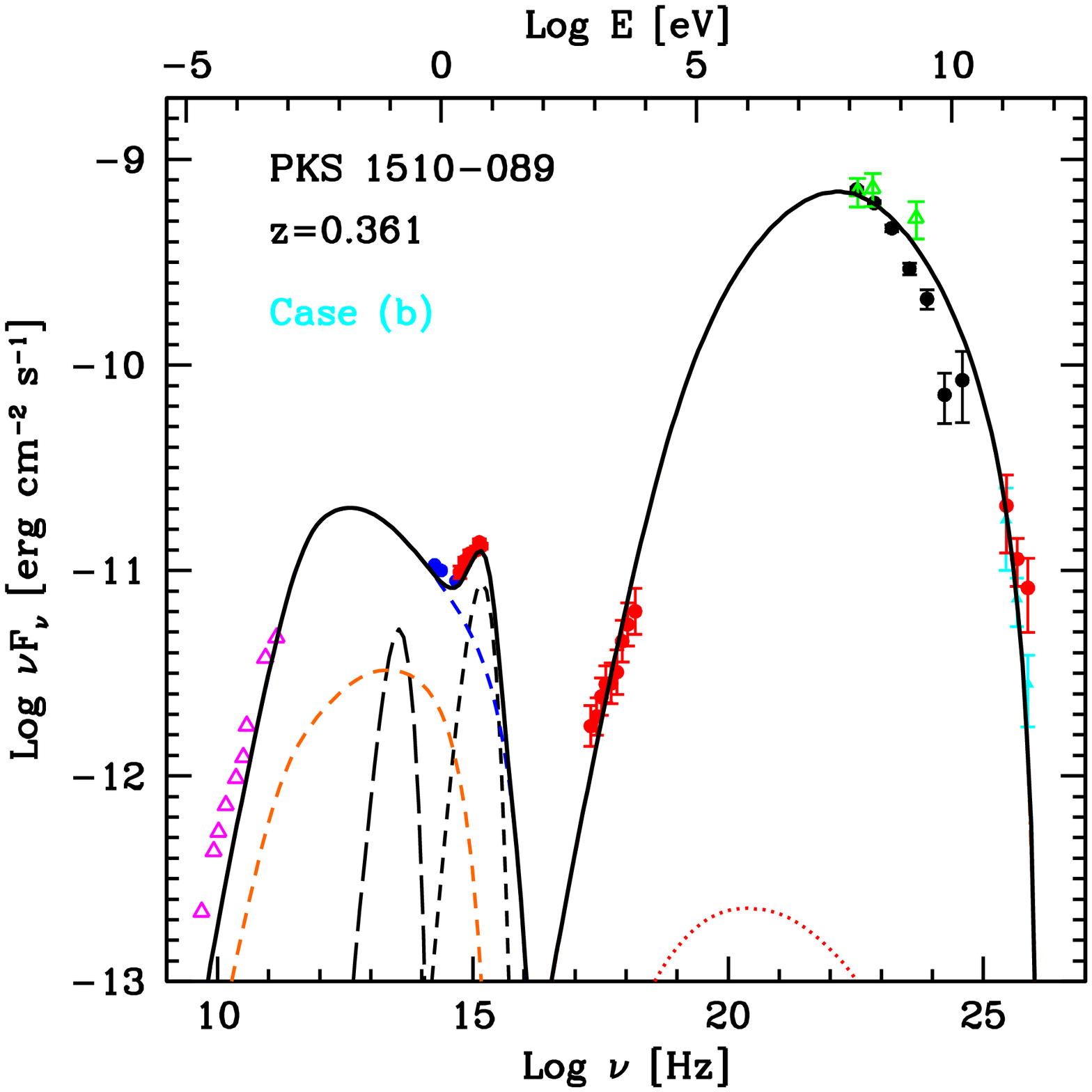}
\caption{SED of PKS~1510$-$089 in 2012 February-April as observed by F-GAMMA and Mets\"ahovi (magenta triangles), GASP-WEBT (blue filled circles), {\it Swift}-UVOT and XRT (red filled circles), {\it Fermi}-LAT (black filled circles), AGILE-GRID (Flare-II, green triangles) and MAGIC (cyan, observed; red, EBL corrected). {\bf Left:} The solid black curve shows the overall emission modelled, where the low energy bump is dominated by the synchrotron emission (blue dashed line) and the high energy bump is dominated by the external Compton mechanism, using the infrared torus (long dashed line) photons as seed photons (case {\it a}). The short dashed line is the thermal component from the accretion disc.
{\bf Right:} The black curve shows the model assuming that the emission region is located at the radio core (case {\it b}). The orange dashed line shows the additional external photon field  representing the slow sheath of the jet. The red dotted line indicates the synchrotron self Compton emission from this region.}
\label{SED}
\end{figure*}

\begin{table*}[th]
\centering
\caption{Model parameters for the two SED models}
\begin{tabular}{lcccccccccccc}
\hline
\hline
model&$\gamma _{\rm min}$ & $\gamma _{\rm b}$ & $\gamma _{\rm max}$ &
$n_1$ & $n_2$ &$B$ & $K$ &$R$ & $\Gamma$ \\
& & & &  & &[G] & [cm$^{-3}]$  & $[10^{16}$
cm]\\ 
\hline
IR torus\tablefootmark{a}&3&9e2&6.5e4&1.9&3.85&0.12&20&30&20\\      
\hline
Sheath\tablefootmark{b}&       800&	7e3&	5e4&	2&	3.4&	1.3e-2&	18&	600& 2.2\\
Spine&	800&	2.6e3&	8e4&	2&	3.7&	6.5e-3&	2.5&	510&	20\\
\hline
\hline
\end{tabular}
\vskip 0.4 true cm
%
\tablefoot{The following quantities are reported: the minimum, break, and maximum Lorentz factors and the low and high energy slope of the electron
energy distribution, 
the magnetic field intensity, the electron density, the radius of the
emitting region and the Lorentz factor.
\tablefoottext{a}{IR torus (external photons for IC scattering provided by the IR torus)}\tablefoottext{b}{Sheath-spine (sheath providing the seed photons for the scattering)}}. 
\label{param}
\end{table*}
 
The observational evidence discussed above (co-spatiality of $\gamma$-ray and millimetre emission and transparency to $>100$\,GeV photons) allow us to locate the emission region outside the BLR but do not provide a clear upper limit for its distance. 
We first tried (case {\it a}) to reproduce the SED finding a solution which minimises the distance from the central engine. The SED is successfully reproduced (Fig.~\ref{SED}) assuming that the emission occurs at a distance of $r\sim1$ pc, i.e.  within the torus indicated with a thick red vertical line in Fig.~\ref{u}. As a consistency check, we infer the run of the optical depth at 37 GHz ($\tau_{37 GHz}$, green line in Fig~\ref{u}) with the distance (for a conical geometry the scaling laws $B\propto r^{-1}$, $K\propto r^{-1}$ can be assumed, and adopting, $B$, $K$ and $R$, of the case {\it a} model), confirming that the  emission region is indeed characterised by $\tau_{37 GHz}<1$ as required by the correlation observed in the light curves. The grey and the yellow areas in Fig.~\ref{u} indicate the opaque and the transparent millimetre regions.

For \object{PKS~1510$-$089} there are measurements of the distance of the VLBA core from the central engine. 
\citet{pushkarev} used the core shift measurements to locate the 15\,GHz VLBA core at $\sim$17.7 parsecs from the central engine. Using the speed and core shift measurement from
that paper gives a distance of $\sim$6.5 parsecs (indicated with a thick cyan line in Fig.~\ref{u}) for the 43\,GHz
core. At this distance, the infrared torus no longer provides 
a strong enough source of seed photons for the IC process
\cite{marscher}
suggest that the dilemma could be solved if the jet is surrounded by
slow sheath providing seed photons for IC scattering. We test this scenario (case {\it b}) by assuming that the
emission blob is surrounded by a sheath with $\Gamma$=2.2, so
that the radiation field is amplified in the emitting region by a Doppler factor $\delta=3.7$ (assuming a viewing angle 2.8$^\circ$ and $\Gamma$=20 for the
blob). 
The fit to the SED is presented in Fig.~\ref{SED}: the orange dashed line represents the observed emission from the modelled sheath, which would be negligible compared to the jet emission and therefore not directly observed. We therefore conclude that, from the point of view of the radiative properties, this scenario is also feasible.

We find that both models provide an acceptable fit to the data and the resulting model parameters are given in Table~\ref{param}. This implies that the presented scenarios are feasible and in agreement with the obtained data; however the parameters used for modelling are not unique. We also note 
that this SED represents an average emission state, since the data have been collected over a few days, and do not account for the rapid ($\sim$ one hour) variability of the $\gamma$-ray emission as measured by the {\it Fermi}-LAT
\citep{saito, 2013arXiv1304.2878F}. As proposed by \citet[][]{marscher} \citep[see also][]{marscher14,narayan}, such rapid flickering could indicate the presence of relativistic turbulent motion within the flow. In this framework, radiation from single, small, turbulent cells is occasionally observed, while the long-term emission is the result of the integrated emission over all of the active jet volume. More detailed simulations along the lines of \citet{marscher14} are required to investigate this scenario in detail. 

\section{Summary and discussion}
In this paper we report the detection of VHE $\gamma$ rays from
\object{PKS~1510$-$089} by the MAGIC telescopes in 2012 February-April. The VHE
$\gamma$-ray flux and spectrum are comparable to those observed from the
source in 2009 March-April by the H.E.S.S. telescopes \citep{hess}. During
the MAGIC observations the source was in a high state in the HE
$\gamma$-ray band, showing significant variability, but the VHE
$\gamma$-ray light curve does not reveal significant variability. This
is in agreement with the result of \cite{hess}. 


We performed a detailed multi-frequency study of the source during
2012 January-April extending for the first time from radio to VHE
$\gamma$ rays. In summary we find that:

\begin{enumerate}
\item The HE and VHE $\gamma$-ray spectra connect smoothly, therefore we conclude that VHE $\gamma$-ray emission and the HE $\gamma$-ray emission originate in a single emission region located outside the BLR. 
\item The VHE $\gamma$-ray observations by MAGIC missed the times of the hour scale variability observed in the HE $\gamma$-ray band and the MAGIC light curve does not show significant variability in daily or weekly time scales. However, the HE $\gamma$-ray variability indicates that within the larger emission region,
there must exist more compact emission regions producing the fast variability.
The model of \citet{marscher14}, in which turbulent plasma flowing at a relativistic speed down the jet and crossing a standing shock, would naturally lead to such behaviour.
We note that, the fast variability could also extend to the VHE $\gamma$-ray band, even if the observations presented here did not detect it.
\item The common variability patterns seen in the HE $\gamma$ ray and 37\,GHz light curves as well as the concurrent ejection of a new component from the 43\,GHz VLBA core support this emission scenario.  
 We also identify several $\sim$180$^\circ$ rotations of the optical polarisation angle, which have been suggested as relating to such events \citep{Marscher_nature}.
\item The SED can be modelled with a one-zone external Compton model for both studied cases, namely: the seed photons originating from the infrared torus and the seed photons originating from a slow sheath of the jet. The latter model is favoured if the VLBA core is as distant from the central engine as suggested by \citet{marscher, pushkarev}.  
\end{enumerate}

However, there are other alternatives for the source of seed photons and for the fast variability:

\begin{itemize}
\item \citet{LeonTavares13} suggested that the relativistic jet could drag the broad
line region clouds to greater distances from the central engine and
the VLBA radio core could be surrounded by such clouds. This would be
manifested by a brightening of the broad emission lines in the
optical monitoring of the spectral lines. We have no such data for our
campaign, but we note that this additional seed photon population is
not required to reproduce our data.

\item It was proposed in \citet{giannios} that additional flickering can be
explained by the jet-in-jet model even if the emission region is far
out in the jet. In this scheme long-term flares are the result of the
``envelope" emission of magnetic reconnection events in the jet, while
short-term flares flag the random formation of ``monster" blobs during
the reconnection process. 
While the discussion in \citet{giannios} was suited for the case of
PKS~1222+216, characterised by shorter time scales (both for the
long-term modulation and the rapid flares), we expect that it might also be possible to reproduce the behaviour of \object{PKS~1510$-$089} by generalising the model.
\end{itemize}

Since \object{PKS~1510$-$089} has been active in $\gamma$ rays in the AGILE and {\it Fermi}-LAT era, there have been several other multi-frequency studies. 
\citet{brown} analysed the {\it Fermi}-LAT data
concluding from the presence of $>20$\,GeV photons that multiple
simultaneously active $\gamma$-ray emission regions are required. We
find that the {\it Fermi}-LAT and MAGIC spectra connect smoothly
suggesting a single emission region. 
These very fast spikes probably originate in a separate
emission region, possibly embedded in the larger region producing the
slower modulations of the radio and $\gamma$-ray light curves. 
Several emission sites were also suggested by \citet{nalewajko12}, who
concluded that the high energy cut-off (in the low state) of the main
synchrotron component implies a two-zone model, otherwise the requested
external photon density is too high. \citet{barnacka} reached a similar conclusion and favoured a two-zone model for reproducing the VHE $\gamma$-ray emission observed by H.E.S.S. telescopes. However, in our modelling, one emission region is sufficient to reproduce the average SED during the high state.

In addition to \object{PKS~1510$-$089}, only two other FSRQs have
been detected in VHE $\gamma$ rays (3C~279 and PKS~1222+216). All
detections have been made during a high state in the lower energy
regimes and even during high activity in the HE $\gamma$-ray band,
3C~279 \citep{aleksic13} and PKS~1222+216 \citep{ackermann14} are detected
in VHE $\gamma$ rays only during individual nights. The upper limits
derived for these two sources from the non-detections are also below
the detected flux. In this sense \object{PKS~1510$-$089} is clearly different
from the other two and follow up observations in lower HE $\gamma$-ray
states should be performed in order to study if the source is a
constant VHE $\gamma$-ray emitter like some of the high peaking BL Lac
objects detected in VHE $\gamma$ rays
\citep[e.g.][]{0414Ver}. However, VHE $\gamma$-ray detections from
FSRQs imply that in all cases the emission takes place outside the BLR \citep{1222, 3C279}. Further observations are needed to study why in some cases we see extremely bright, fast flares of VHE $\gamma$ rays and in other cases (as in the case of \object{PKS~1510$-$089}) the emission of VHE $\gamma$ rays appears more stable.

\begin{acknowledgements}
We would like to thank the Instituto de Astrof\'{\i}sica de
Canarias for the excellent working conditions at the
Observatorio del Roque de los Muchachos in La Palma.
The support of the German BMBF and MPG, the Italian INFN, 
the Swiss National Fund SNF, and the Spanish MICINN is 
gratefully acknowledged. This work was also supported by the CPAN CSD2007-00042 and MultiDark
CSD2009-00064 projects of the Spanish Consolider-Ingenio 2010
programme, by grant DO02-353 of the Bulgarian NSF, by grant 127740 of 
the Academy of Finland,
by the DFG Cluster of Excellence ``Origin and Structure of the 
Universe'', by the DFG Collaborative Research Centres SFB823/C4 and SFB876/C3,
by the Polish MNiSzW grant 745/N-HESS-MAGIC/2010/0 and by JSPS KAKENHI Grants numbers 24000004 and 25800105.

The \textit{Fermi}-LAT Collaboration acknowledges generous ongoing support
from a number of agencies and institutes that have supported both the
development and the operation of the LAT as well as scientific data
analysis. These include the National Aeronautics and Space Administration
and the Department of Energy in the United States, the Commissariat \`a
l'Energie Atomique and the Centre National de la Recherche Scientifique /
Institut National de Physique Nucl\'eaire et de Physique des Particules in
France, the Agenzia Spaziale Italiana and the Istituto Nazionale di Fisica
Nucleare in Italy, the Ministry of Education, Culture, Sports, Science and
Technology (MEXT), High Energy Accelerator Research Organization (KEK) and
Japan Aerospace Exploration Agency (JAXA) in Japan, and the
K.~A.~Wallenberg Foundation, the Swedish Research Council and the
Swedish National Space Board in Sweden.

Additional support for science analysis during the operations phase is
gratefully acknowledged from the Istituto Nazionale di Astrofisica in
Italy and the Centre National d'\'Etudes Spatiales in France.

Astrorivelatore Gamma ad Immagini LEggero (AGILE) is a scientific mission
of the Italian Space Agency (ASI) with INFN, INAF e CIFS participation.
AGILE research partially supported through the ASI grants 
I/089/06/2, I/042/10/0 and I/028/12/0.

Data from the Steward Observatory spectropolarimetric monitoring project were used. This programme is supported by Fermi Guest Investigator grants NNX08AW56G, NNX09AU10G, and NNX12AO93G.

This article is partly based on observations made with the telescopes IAC80
and TCS operated by the Instituto de Astrofisica de Canarias in the Spanish
Observatorio del Teide on the island of Tenerife. Most of the observations
were taken under the rutinary observation programme. The IAC team
acknowledges the support from the group of support astronomers and
telescope operators of the Observatorio del Teide.

The Abastumani team acknowledges financial support of the project FR/638/6-320/12 by the Shota Rustaveli National Science Foundation under contract 31/77.

The OVRO 40-m monitoring programme is
supported in part by NASA grants NNX08AW31G 
and NNX11A043G, and NSF grants AST-0808050 
and AST-1109911.

The VLBA is operated by the National Radio Astronomy Observatory. The
National Radio Astronomy Observatory is a facility of the National
Science Foundation operated under cooperative agreement by Associated
Universities, Inc. The research at Boston University (BU) was funded
in part by NASA Fermi Guest Investigator grants NNX11AQ03G,
NNX11AO37G, and NNX12AO90G. The PRISM camera at Lowell Observatory was
developed by K.\ Janes et al. at BU and Lowell Observatory, with
funding from the NSF, BU, and Lowell Observatory.

The Submillimeter Array is a joint project between the Smithsonian
Astrophysical Observatory and the Academia Sinica Institute of Astronomy
and Astrophysics and is funded by the Smithsonian Institution and the
Academia Sinica.

The Mets\"ahovi team acknowledges the support from the Academy of Finland
to our observing projects (numbers 212656, 210338, 121148, and others).

This research is partly based on observations with the 100-m telescope of
the MPIfR (Max-Planck-Institut f\"ur Radioastronomie) at Effelsberg and with the IRAM 30-m
telescope. IRAM is supported by INSU/CNRS (France), MPG (Germany) and IGN
(Spain). I. Nestoras is funded by the International Max Planck Research
School (IMPRS) for Astronomy and Astrophysics at the Universities of Bonn and Cologne.

\end{acknowledgements}


\begin{thebibliography}{}
\bibitem[Abdo et al.(2010a)]{abdo10}
Abdo, A. A., Ackermann, M., Agudo, I. et al. ({\it Fermi}-LAT Collaboration) 2010a, ApJ, 721, 1425
\bibitem[Abdo et al.(2010b)]{nature} 
Abdo, A. A., Ackermann, M., Ajello, M. et al. ({\it Fermi}-LAT Collaboration) 2010b, Nature, 463, 919
\bibitem[Abramowski et al.(2013)]{hess}
Abramowski, A., Acero, F., Aharonian, F. et al. (H.E.S.S. Collaboration) 2013, A\&A, 554, 107
\bibitem[Ackermann et al.(2012)]{ackermann12} 
Ackermann, M., Ajello, M., Allafort, A., et al. ({\it Fermi}-LAT Collaboration) 2012, ApJ, 747, 104
\bibitem[Ackermann et al.(2014)]{ackermann14} 
Ackermann, M., Ajello, M., Allafort, A., et al. ({\it Fermi}-LAT and MAGIC Collaborations), 2014, ApJ, 786, 157
\bibitem[Albert et al.(2007)]{Albert07} Albert, J., Aliu, E., Anderhub, H., et al. (MAGIC Collaboration) 2007, \apj, 669, 862 
\bibitem[Albert et al.(2008a)]{Science} 
Albert, J., Aliu, E., Anderhub, H. et al. (MAGIC Collaboration) 2008, Science, 320, 1752
\bibitem[Albert et al.(2008b)]{Albert08_RF} Albert, J., Aliu, E., Anderhub, H., et al. (MAGIC Collaboration) 2008b, Nucl. Instr. Meth. A, 588, 424 
\bibitem[Aliu et al.(2012)]{0414Ver}
Aliu, E., Archambault, S., Arlen, T. (VERITAS Collaboration) et al. 2012, ApJ, 755, 118
\bibitem[Aleksi\'c et al.(2010)]{Aleksic10} Aleksi\'c, J., Anderhub, H., Antonelli, L.~A., et al. (MAGIC Collaboration) 2010, A\&A, 519, 32 
\bibitem[Aleksi\'c et al.(2011a)]{1222} 
Aleksi\'c, J., Anderhub, H., Antonelli, L.~A., et al.  (MAGIC Collaboration) 2011a, ApJ, 730, L8 
\bibitem[Aleksi\'c et al.(2011b)]{3C279}
Aleksi\'c, J., Anderhub, H., Antonelli, L.~A., et al. (MAGIC Collaboration) 2011b, A\&A, 530, A4
\bibitem[Aleksi\'c et al.(2012)]{Performance}
Aleksi\'c, J., Alvarez, E. A., Antonelli, L. A., et al. (MAGIC Collaboration) 2012, Astropart. Phys, 35, 435
\bibitem[Aleksi\'c et al.(2013)]{2344}
Aleksi\'c, J., Antonelli, L. A., Antoranz, P., et al. (MAGIC Collaboration) 2013, A\&A, 556, A67
\bibitem[Aleksi\'c et al.(2014)]{aleksic13}
Aleksi\'c, J., Ansoldi, S., Antonelli, L. A., et al.(MAGIC Collaboration) 2014, accepted to A\&A
\bibitem[Aliu et al.(2009)]{Aliu09} Aliu, E., Anderhub, H., Antonelli, L.~A., et al. (MAGIC Collaboration) 2009, ApJL, 692, L29  
\bibitem[Angelakis et al.(2008)]{2008arXiv0809.3912A} Angelakis, E.,
Fuhrmann, L., Marchili, N., Krichbaum, T.~P., \& Zensus, J.~A.\ 2008, arXiv:0809.3912
\bibitem[Arnaud (1996)]{arnaud96} Arnaud, K. A. 1996, ASPC, 101, 17 
\bibitem[Atwood et al.(2009)]{atwood09}
Atwood, W.~B., Abdo, A. A., Ackermann, M., et al. 2009, \apj, 697, 1071
\bibitem[Baars et al.(1977)]{Baars77} Baars, J. W. M., Genzel, R., Pauliny-Toth, I. I. K., Witzel, A., 1977, A\&A, 61, 1, 99
\bibitem[Barnacka et al.(2013)]{barnacka}Barnacka, A., Moderski, R., Behera, B., Brun, P. \& Wagner, S., 2013, submitted to A\&A, arXiv:1307.1779
\bibitem[Barthelmy et al.(2005)]{barthelmy05} Barthelmy, S. D., Barbier, L. M., Cummings, J. R. et al. 2005, SSRv, 120, 143
\bibitem[Beaklini et al.(2011)]{brazil-atel} Beaklini, P., Abraham Z., Dominici, T. 2011, The Astronomer's Telegram 3799
\bibitem[Bertero(1989)]{Bertero} Bertero, M., 1989, Advances in Electronics and Electron Physics, 75, Academic Press Inc., New York
\bibitem[Brown et al.(2013)]{brown} Brown, A. M. 2013, MNRAS 431,824
\bibitem[Bulgarelli et al.(2012)]{bulgarelli} Bulgarelli, A., Chen, A. W., Tavani, M. et al., 2012 A\&A 540, A79
\bibitem[Burrows et al.(2005)]{burrows05} Burrows, D. N., Hill, J. E., Nousek, J. A. et al. 2005, SSRv, 120, 165  
\bibitem[B\"ottcher et al.(2009)]{bottcher}
B\"ottcher M., Reimer, A. \& Marscher, A. P. 2009, ApJ, 703, 1168
\bibitem[Cardelli et al.(1989)]{cardelli}
Cardelli, J. A.,  Clayton, G. C., Mathis, J. S.	1989, ApJ, 345, 245 
\bibitem[Cash (1979)]{cash79} Cash, W. 1979, ApJ, 228, 939
\bibitem[Celotti et al.(1997)]{celotti} Celotti, A., Padovani, P., Ghisellini, G. 1997, MNRAS, 286, 415
\bibitem[Chatterjee et al.(2008)]{chatterjee08} Chatterjee, R., Jorstad, S. G., Marscher, A. P. et al. 2008, ApJ, 689, 79
\bibitem[Chincarini et al.(2003)]{Chi03} Chincarini, G., Zerbi, F. M., Antonelli, L. A., et~al.\ 2003, The Messenger, 113, 40
\bibitem[Cortina(2012)]{atel} 
Cortina (on behalf of the MAGIC collaboration) 2012 The Astronomer's Telegram 3975
\bibitem[Covino et al.(2004a)]{Cov04a} Covino, S., Zerbi, F. M., Chincarini, G., et~al.\ 2004a, AN, 325, 543
\bibitem[Covino et al.(2004b)]{Cov04b} Covino, S., Stefanon, M., Sciuto, G., et~al.\ 2004b, SPIE, 5492, 1613
\bibitem[D'Ammando et al.(2009)]{atel1957} D'Ammando, F., Vercellone, S.,  Tavani, M. et al. 2009, The Astronomer's Telegram 1957
\bibitem[D'Ammando et al.(2011)]{dammando11} D'Ammando, F., Raiteri, C. M., Villata, M. et al. 2011, A\&A, 529, 145
\bibitem[De Caneva et al.(2012)]{decaneva} 
De Caneva, G., Barres de Almeida, U. and Becerra Gonzelez, J. et al. (on behalf of the MAGIC collaboration) 2012, proceedings of the''5th International Symposium on High-Energy Gamma-Ray Astronomy”, AIP Conf.\ Proc.\ {\bf 1505}, 502 
\bibitem[Devillard(1997)]{Dev97} Devillard, N. 1997, The Messenger, 87
\bibitem[Dermer \& Schlickeiser(1994)]{dermer} Dermer, C. \& Schlickeiser, R., 1994, ApJS, 90, 945
\bibitem[Dominguez, et al.(2011)]{dominguez}
Dominguez, A., Primack, J. R., Rosario, D. J. et al. 2011, MNRAS, 410, 2556
\bibitem[Donea \& Protheroe(2003)]{donea} Donea \& Protheroe 2003, APh, 18, 377 
\bibitem[Donnarumma, et al.(2011)]{atel3470}
Donnarumma, I., Lucarelli, F., Vercellone, S. et al. 2011, The Astronomer's Telegram 3470
\bibitem[Fomin et al.(1994)]{Fomin94} Fomin, V.~P., Stepanian, A.~A., Lamb, R.~C., et al. 1994, Astropart. Phys., 2, 137
\bibitem[Foschini et al.(2013)]{2013arXiv1304.2878F} Foschini, L.,
Bonnoli, G., Ghisellini, G., et al.\ 2013, A\&A, 555, 138
\bibitem[Franceschini et al.(2008)]{Fra08} Franceschini, A., Rodighiero, G., \& Vaccari, M. 2008, \aap, 487, 837
\bibitem[Fuhrmann et al.(2007)]{2007AIPC..921..249F} Fuhrmann, L., Zensus,
J.~A., Krichbaum, T.~P., Angelakis, E., \& Readhead, A.~C.~S.\ 2007, The First GLAST Symposium, 921, 249
\bibitem[Fuhrmann et al.(2008)]{2008A&A...490.1019F} Fuhrmann, L., Krichbaum, T. P., Witzel, A., et al. 2008, \aap, 490, 1019
\bibitem[Gehrels et al.(2004)]{gehrels04} Gehrels, N., Chincarini, G., Giommi, P. et al. 2004, ApJ, 611, 1005
\bibitem[Ghisellini \& Tavecchio(2009)]{ghisellini09}
Ghisellini, G. et al. 1998, MNRAS, 301, 451
\bibitem[Giannios(2013)]{giannios} Giannios, D.\ 2013, \mnras, 431, 355
\bibitem[Giommi et al.(2012)]{giommi2012} Giommi, P., Polenta, G., L{\"a}hteenm{\"a}ki, A., et al.\ 2012, \aap, 541, A160
\bibitem[Giuliani et al.(2004)]{Giuliani2004:diff_model}
Giuliani, A., Chen, A., Mereghetti, S. et al. 2004, Mem.\ SAIt Suppl. 5, 135 
\bibitem[Gurwell et al.(2007)]{gurwell07} Gurwell, M.~A., Peck, A.~B.,
Hostler, S.~R., et~al. 2007, Astronomical Society of the Pacific
Conference Series, 375, 234
\bibitem[G{\'o}rski et al.(2005)]{2005ApJ...622..759G}
G{\'o}rski, K.~M., Hivon, E., Banday, A.~J. et al. 2005, ApJ, 622, 759  
\bibitem[Hartman et al.(1999)]{hartman99}
Hartman, R. C., Bertsch, D. L., Bloom, S. D. et al. 1999, ApJS..123...79
\bibitem[Hartman et al.(2000)]{hartman} Hartman, R. C., B\"ottcher, M., Aldering, G. et al. 2001, ApJ, 553, 683
\bibitem[Hauser \& Dwek(2001)]{Hau01} Hauser, M.~G., \& Dwek, E. 2001, \araa, 39, 249
\bibitem[Jorstad et al.(2001)]{jorstad01} Jorstad, S. G., Marscher, A. P., Mattox, J. R. et al. 2001, ApJ, 556, 738
\bibitem[Jorstad et al.(2005)]{jorstad05} Jorstad, S. G., Marscher, A. P., Lister, M. L. et al. 2005, AJ, 130, 1418
\bibitem[Jorstad et al.(2007)]{jorstad07}
Jorstad, S. G., Marscher, A. P., Stevens, J. A. et al. 2007, AJ, 134, 799
\bibitem[Jorstad et al.(2012)]{jorstad11} Jorstad, S. G., Marscher, A. P., Agudo, I. et al. 2011, AAS Meeting \#217, \#408.04; Bulletin of the American Astronomical Society, Vol. 43, 2011 
\bibitem[Jorstad et al.(2013)]{jorstad13}  Jorstad, S. G., Marscher, A. P. et al. 2013, ApJ, 773, 147
\bibitem[Kalberla et al.(2005)]{kalberla05} Kalberla, P. M. W., Burton, W. B., Hartmann, D. al. 2005, A\&A, 440, 775
\bibitem[Kaspi et al.(2000)]{kaspi}
Kaspi, S., Smith, P. S., Netzer, H., et al. 2000, ApJ, 533, 631
\bibitem[Kataoka et al.(2008)]{kataoka08} Kataoka, J., Madejski, G., Sikora, M. et al. 2008, ApJ, 672, 787
\bibitem[Krimm et al.(2013)]{krimm} Krimm, H. A., Holland, S. T., Corbet, R. H. D. et al. 2013, ApJS, 209, 14
\bibitem[Larionov et al.(2008)]{larionov}
Larionov, V. Jorstad, S. G., Marscher, A. P., et al. (2008), A\&A, 492, 389
\bibitem[Leon-Tavares et al.(2013)]{LeonTavares13}
Leon-Tavares, J., Chavushyan, V., Patino-Alvarez, V. et al. 2013, ApJ, 763, 36 
\bibitem[Li \& Ma(1983)]{Li83} Li, T.~P., \& Ma, Y.~Q. 1983, \apj, 272, 317
\bibitem[Lindfors et al.(2005)]{lindfors05}
Lindfors, E., Valtaoja, E., T\"urler, M. 2005, A\&A, 440, 845
\bibitem[Lindfors et al.(2006)]{lindfors06}
Lindfors, E., T\"urler, M., Valtaoja, E. et al. 2006, A\&A, 456, 895
\bibitem[Lindfors et al.(2013)]{fermisymp}
Lindfors, E., Nilsson, K., Barres de Almeida, U. et al. (on behalf of the MAGIC collaboration), 2012 Fermi Symposium proceedings - eConf C121028 [arXiv:1303.2102] 
\bibitem[Lucarelli et al.(2012)]{lucarelli} Lucarelli, F., Piano, G., Verrecchia, F. et al. 2012, The Astronomer's Telegram 3934
\bibitem[L\"ahteenm\"aki \& Valtaoja, E. (2003)]{lahteenmaki} 
L\"ahteenm\"aki, A. \& Valtaoja, E., 2003, ApJ, 590, 92
\bibitem[Malmrose et al. (2011)]{malmrose} 
Malmrose, M. P., Marscher, A. P., Jorstad, S. G., Nikutta, R., Elitzur, M. ApJ, 732, 116
\bibitem[Maraschi et al.(1992)]{maraschi92} Maraschi, L., Ghisellini, G. \& Celotti, A. 1992, ApJ, 397, 5
\bibitem[Maraschi et al.(2003)]{maraschi03} Maraschi, L., \& Tavecchio, F. 2003, ApJ, 593, 667
\bibitem[Marscher \& Gear(1985)]{MG85} Marscher, A. P. \& Gear, W. K., 1985, ApJ, 298, 114
\bibitem[Marscher et al.(2008)]{Marscher_nature}
Marscher, A., Jorstad, S., D'Arcangelo, F. D. et al. 2008, Nature,452,966
\bibitem[Marscher et al.(2010)]{marscher}
Marscher, A., Jorstad, S., Larionov, V. et al. 2010, ApJL, 710, L126
\bibitem[Marscher et al.(2012)]{marscher12}
Marscher, A., Jorstad, S., Agudo, I. et al. 2012, Fermi \&Jansky Proceedings -eConf C1111101[arXiv:1204.6707]
\bibitem[Marscher(2014)]{marscher14}
Marscher, A. 2014, ApJ, 780, 87
\bibitem[Mattox et al.(1996)]{mattox96}
Mattox, J.~R., Bertsch, D. L., Chiang, J. et al. 1996, \apj, 461, 396
\bibitem[Moralejo et al.(2009)]{Moralejo09} Moralejo, A., Gaug, M., Carmona, E., et al. 2009, in Proc. 31st ICRC ({\L}\'od\'z), 469 (arXiv:0907.0943)
\bibitem[Naghizadeh-Khouei \& Clarke(1993)]{Naghizadeh-Khouei}
Naghizadeh-Khouei \& Clarke, 1993, A\&A, 274, 968
\bibitem[Nalewajko et al.(2010)]{nalewajko10}
Nalewajko, K. 2010, International Journal of Modern Physics D, 19, 701
\bibitem[Nalewajko et al.(2012)]{nalewajko12}
Nalewajko, K., Sikora, M., Madejeski, G. et al. 2012, ApJ, 760, 69
\bibitem[Narayan \& Piran(2012)]{narayan}	
Narayan, R. \& Piran, T. 2012, MNRAS, 420, 1, 604
\bibitem[Nolan et al.(2012)]{nolan}
Nolan, P. L., Abdo, A. A., Ackermann, M. et al. 2012, ApJS, 199, 31
\bibitem[Nestoras et al.(2011)]{fgamma-atel} Nestoras, I., Fuhrmann, L., Angelakis, E. et al. 2011, The Astronomer's Telegram 3698 
\bibitem[Orienti et al.(2011)]{medicina-atel} Orienti, M., D'Ammando, F., Giroletti M. et al. 2011, The Astronomer's Telegram 3775 
\bibitem[Orienti et al.(2013)]{orienti} 
Orienti, M., D'Ammando, F., Giroletti, M. et al. 2013, MNRAS, 428, 241
\bibitem[Pacciani et al.(2012)]{pacciani} Pacciani, L., Donnarumma, I., Denney, K. D. et al. 2012, MNRAS, 425, 2015
\bibitem[Pittori et al.(2009)]{Pittori2009:Catalogue}
Pittori, C., Verrecchia, F., Chen, A.~W. et al. 2009 \aap, 506, 1563
\bibitem[Poole et al.(2008)]{poole} Poole, T. S., Breeveld, A. A., Page, M. J. et al. 2008, MNRAS, 383, 62
\bibitem[Poutanen \& Stern(2010)]{PoutanenStern}
Poutanen, J. \& Stern, B. 2010, ApJ, 717, L118
\bibitem[Pucella et al.(2008)]{Pucella08}
Pucella, G., Vittorini, V., D'Ammando F., et al. 2008, \aap, 491, L21
\bibitem[Pushkarev et al.(2012)]{pushkarev} Pushkarev, A. B., Hovatta, T., Kovalev, Y. Y. et al. 2012 A\&A, 545, 113
\bibitem[Raiteri et al.(1998)]{raiteri}
Raiteri, C. M., Villata, M., de Francesco, G. et al. 1998, A\&AS, 130, 495
\bibitem[Raiteri et al.(2008)]{raiteri08}
Raiteri, C. M., Villata, M., Larionov, V. M. et al. 2008, A\&A, 491, 755
\bibitem[Raiteri et al.(2010)]{2010A&A...524A..43R} Raiteri, C.~M., Villata, M., Bruschini, L., et al.\ 2010, \aap, 524, A43
 \bibitem[Raiteri et al.(2012)]{raiteri12}
Raiteri, C. M., Villata, M. Smith, P. S. et al. 2012, A\&A, 545, A48
\bibitem[Roming et al.(2005)]{UVOT} Roming, P. W. A., Kennedy, T. E., Mason, K. O. et al. 2005, SSRv, 120, 95
\bibitem[Richards et al.(2011)]{Richards11} Richards, J. L., Max-Moerbeck, W., Pavlidou, V. et al. 2011, ApJS, 194, 29
\bibitem[Saito et al.(2013)]{saito} Saito, S., Stawarz, L., Tanaka, Y. et al. 2013, ApJ, 766, 11 
\bibitem[Savolainen et al.(2002)]{savolainen}
Savolainen, T., Wiik, K., Valtaoja, E. et al. 2002, A\&A, 394, 851
\bibitem[Schlegel et al.(1998)]{schlegel}
Schlegel, D. J., Finkbeiner, D. P., Davis, M. et al. 1998, ApJ, 500, 525 
\bibitem[Sikora et al.(1994)]{sikora94} Sikora, M., Begelman, M. C., Rees, M. J. 1994, ApJ, 421, 153
\bibitem[Sikora et al.(2008)]{sikora08} Sikora, M., Moderski, R., Madejski, G. M. 2008, ApJ, 675, 71 
\bibitem[Sikora et al.(2009)]{sikora09} Sikora, M., Stawarz, L., Moderski, R. Nalewajko, K. \& Madejski, G. M. 2009, ApJ, 704, 38
\bibitem[Sitarek  \& Bednarek(2008)]{sitarek} Sitarek, J., \& Bednarek, W., 2008, MNRAS, 391, 624
\bibitem[Sitarek et al.(2013)]{2013arXiv1305.1007S} Sitarek, J., Gaug,
M., Mazin, D., Paoletti, R., \& Tescaro, D. 2013, arXiv:1305.1007
\bibitem[Smith et al.(2009)]{smith} Smith, P.S., Montiel, E., Rightley, S., Turner, J., Schmidt, G.D., \& Jannuzi, B.T. 2009, arXiv:0912.3621, 2009 Fermi Symposium, eConf Proceedings C091122
\bibitem[Stecker et al.(1992)]{Ste92} Stecker, F.~W., de Jager, O.~C., \& Salamon, M.~H. 1992, \apj, 390, L49
\bibitem[Striani et al.(2010)]{atel2385}
Striani, E., Verrecchia F., Tavani M. et al. 2010, The Astronomer's Telegram 2385
\bibitem[Tavani et al.(2009)]{Tavani2009:AGILE}
Tavani, M., Barbiellini, G., Argan, A. et al. 2009,\aap, 502, 995     
\bibitem[Tavecchio et al.(1998)]{tavecchio98}
Tavecchio, F., Maraschi, L., \& Ghisellini, G. 1998, ApJ, 509, 608
\bibitem[Tavecchio \& Mazin(2009)]{FM}
Tavecchio, F.\& Mazin, D. 2009, MNRAS, 392, 40
\bibitem[Tavecchio et al.(2013)]{tavecchio13} Tavecchio, F., Pacciani, L., Donnarumma, I. et al. 2013, MNRAS, 435,24
\bibitem[Ter\"asranta et al.(1998)]{Metsaehovi} Teraesranta, H., Tornikoski, M., Mujunen, A. et al. 1998, A\&AS, 132, 305
\bibitem[Vercellone et al.(2010)]{vercellone}
Vercellone, S., D'Ammando, F., Vittorini, V. et al. 2010, ApJ, 712, 405 
\bibitem[Verrecchia et al.(2012)]{ATel3907}
Verrecchia, F., Pittori, C. Cardillo, M. et al. 2012, The Astronomer's Telegram 3907 
\bibitem[Verrecchia et al. (2013)]{verrecchia13} 
Verrecchia F., Pittori C., Chen A. et al., 2013 A\&A, 558, 137 
\bibitem[Villata et al.(1997)]{villata97}
Villata M., Raiteri, C. M., Ghisellini, G. et al. 1997 A\&AS 121, 119
\bibitem[Wagner(2010)]{wagner}
Wagner (H.E.S.S. Collaboration) HEAD meeting \#11, \#7.05; Bulletin of the American Astronomical Society, Vol. 41, p.660
\bibitem[Zerbi et al.(2001)]{Zer01} 
Zerbi, F. M., Chincarini, G., Ghisellini, G., et~al.\ 2001, AN, 322, 275

\end{thebibliography}
\end{document}